\documentclass[twocolumn,prb,floatfix,superscriptaddress,nobibnotes, maxnames=15]{revtex4-2}
\usepackage{graphicx}
\usepackage{enumerate}
\usepackage[colorlinks=true,citecolor=blue]{hyperref}
\usepackage{textcomp}
\usepackage{amsmath}
\usepackage{amssymb}
\usepackage{pgf}
\usepackage{subfigure}
\usepackage[english]{babel}
\usepackage[normalem]{ulem}
\usepackage{tabularray}
\usepackage{xcolor}
\usepackage{array}

\def\ket#1{| #1\rangle}

\newcommand{\ua}{\uparrow}
\newcommand{\da}{\downarrow}

\graphicspath{{./}{./Figures/}}

\begin{document}
\title{Probing Majorana localization in minimal Kitaev chains through a quantum dot}

\author{Rub\'en Seoane Souto}
\affiliation{Division of Solid State Physics and NanoLund, Lund University, S-22100 Lund, Sweden}
\affiliation{Center for Quantum Devices, Niels Bohr Institute, University of Copenhagen, 2100 Copenhagen, Denmark}
\affiliation{Departamento de Física Teórica de la Materia Condensada, Condensed Matter Physics Center (IFIMAC) and Instituto Nicolás Cabrera, Universidad Autónoma de Madrid, 28049 Madrid, Spain}
\affiliation{Instituto de Ciencia de Materiales de Madrid (ICMM), Consejo Superior de Investigaciones Científicas (CSIC),
Sor Juana Inés de la Cruz 3, 28049 Madrid, Spain}

\author{Athanasios Tsintzis}
\affiliation{Division of Solid State Physics and NanoLund, Lund University, S-22100 Lund, Sweden}

\author{Martin Leijnse}
\affiliation{Division of Solid State Physics and NanoLund, Lund University, S-22100 Lund, Sweden}
\affiliation{Center for Quantum Devices, Niels Bohr Institute, University of Copenhagen, 2100 Copenhagen, Denmark}

\author{Jeroen Danon}
\affiliation{Department of Physics, Norwegian University of Science and Technology, 7491 Trondheim, Norway}

\begin{abstract}
Artificial Kitaev chains, formed by quantum dots coupled via superconductors, have emerged as a promising platform for realizing Majorana bound states.
Even a minimal Kitaev chain (a quantum dot--superconductor--quantum dot setup) 
can host Majorana states at discrete sweet spots.
However, unambiguously identifying Majorana sweet spots in such a system is still challenging. In this work, we propose an additional dot coupled to one side of the chain as a tool to identify good sweet spots in minimal Kitaev chains.
When the two Majorana states in the chain overlap, the extra dot couples to both and thus splits an even--odd ground-state degeneracy when its level is on resonance.
In contrast, a ground-state degeneracy will persist for well-separated Majorana states. This difference can be used to identify points in parameter space with spatially separated Majorana states, using tunneling spectroscopy measurements. We perform a systematic analysis of different relevant situations. We show that the additional dot can help distinguishing between Majorana sweet spots and other trivial zero-energy crossings. We also characterize the different conductance patterns, which can serve as a guide for future experiments aiming to study Majorana states in minimal Kitaev chains.
\end{abstract}

\maketitle

\section{Introduction}

Majorana bound states that emerge at the ends of one-dimensional topological superconductors~\cite{Kitaev_2001} have attracted significant attention due to their exotic physical properties and their potential applications in quantum technologies~\cite{Leijnse_Review2012,Alicea_RPP2012,beenakker2013search,Aguado_Nuovo2017,BeenakkerReview_20,flensberg2021engineered,Marra_Review2022}. Superconductor--semiconductor nanowires have been proposed as a platform for engineering these states~\cite{Oreg_PRL2010,Lutchyn_PRL2010} and experimental signatures consistent with their existence have been observed~\cite{Lutchyn_NatRev2018}, relying on transport measurement of zero-energy states~\cite{Mourik_science2012,Nichele_PRL2017,Fornieri_Nat2019,Banerjee_PRB2023}, non-local conductance~\cite{Aghaee_PRB2023_short,Banerjee_PRL2023}, and state localization~\cite{Albrecht_Nature2016}. However, these observations can also be mimicked by trivial states originating from alternative mechanisms, such as disorder and smooth confining potentials~\cite{Prada_PRB2012, Kells_PRB12, Liu2012, Liu2017, Moore_PRB18,reeg2018zero,Awoga_PRL2019,Vuik_SciPost19,Pan_PRR20,Prada_review,hess2021local}. Thus, a key challenge in the field is to mitigate the effects of disorder.

In recent years, arrays of quantum dots coupled to nanoscale superconductors have emerged as a promising alternative for probing Majorana physics, thanks to their robustness to disorder in the material~\cite{Sau_NatComm2012}. In the simplest configuration, two quantum dots couple to a single superconductor that facilitates crossed Andreev reflection (CAR). This setup was previously studied in the context of Cooper pair splitters~\cite{Recher_PRB2001,Hofstetter_Nature2009,Herrmann_PRL2010,Fulop_PRL2015,Wang_Nat2022,Bordoloi_Nat2022}. The central superconductor also allows the transference of single electrons via elastic cotunneling (ECT). This setup, a so-called minimal Kitaev chain, can host ``poor man's Majorana states'' (PMMs) at specific gate configurations~\cite{Leijnse_PRB2012}. While PMMs are not topologically protected, they share certain topological properties with genuine topological Majoranas, including non-abelian characteristics~\cite{tsintzis2023roadmap,Boross_2023}.

Initial experiments have successfully demonstrated the existence of ``sweet spots'' where the measured local and non-local conductance align with theoretical predictions for PMMs~\cite{Leijnse_PRB2012,Liu_PRL2022,Dvir2023,Bordin_arXiv2023}. While the basic physics of a single PMM system captured in a toy model agrees with experimental findings, real experiments involve additional complexities, such as interactions, finite Zeeman splitting, excited states, and potentially strong coupling between the dots and the central superconductor~\cite{Tsintzis2022}. These additional factors make the overall picture more intricate, leading to low-quality sweet spots, i.e., zero-energy crossing with overlapping Majorana states. The Majorana polarization (MP) provides a measurement for the Majorana localization~\cite{Sedlmayr2015, Sedlmayr2016,Aksenov2020}, and therefore, the quality of the sweet spots. The unambiguous identification of sweet spots with high MP (good Majorana localization) based solely on transport measurements remains a challenge. This necessitates the development of alternative approaches to identify high-MP PMM regimes before embarking on more complex experiments involving topological qubits, fusion, or braiding~\cite{tsintzis2023roadmap,Boross_2023,Liu2022_PMMfusion}.

In this work, we investigate the idea to use an additional dot to measure the Majorana quality and localization. A coupling between two Majorana states splits the ground state degeneracy associated with the fermionic mode they define. However, it is possible to find regimes where two spatially overlapping Majoranas remain uncoupled and thus at (almost) zero energy, see for example~\cite{Prada_PRB2012, Kells_PRB12, Liu2012}. An additional quantum dot coupled to one side of the system can then mediate a coupling between the Majoranas, lifting the ground state degeneracy.
This mechanism thus provides a way of testing the Majorana localization at the end of nanowires~\cite{deng2016majorana,Prada_PRB2017,Clarke_PRB2017,deng2018nonlocality,gruñeiro2023transport}. When the energy of the dot level is swept, local conductance spectroscopy shows typically a ``diamond" or ``bowtie" pattern, depending on whether the tunneling to the dot or the direct coupling between the Majorana states dominates~\cite{Prada_PRB2012}.
In contrast, well-separated Majorana states show up as persistent zero-energy conductance peaks, independently from the on-site energy of and tunnel coupling to the additional dot.

Here, we take inspiration from these previous works on proximitized nanowires and extend the study to minimal Kitaev chains, where all relevant parameters are tunable through electrostatic gates, see also \cite{tsintzis2023roadmap}. We find similar physics to the nanowire case, suggesting that spectroscopy through an additional quantum dot is indeed a method to identify high-MP sweet spots. By tuning the energy of the PMM dots, we find different patterns, including the predicted bowtie and diamond structures, thus providing a way to characterize the Majorana overlap in the system.
Using a simple toy model for the combined dot--PMM system, we develop an analytic understanding of the structure of these patterns, providing also a deeper insight into the physics underlying the quantum dot-Majorana wire coupling, studied before~\cite{deng2016majorana,Prada_PRB2017,Clarke_PRB2017,deng2018nonlocality,gruñeiro2023transport}.

The article is structured as follows. First, we introduce the model Hamiltonian, in Sec.~\ref{sec:model}. We then present the toy model and explore the physics of the single quantum dot coupled to a minimal Kitaev chain, where we disregard the spin degree of freedom and ignore interactions.
In Sec.~\ref{sec:anal} we use the toy model to get deeper analytic insight in the mechanisms underlying the so-called diamond and bowtie patterns in the level structure.
At the end of the article, in Sec.~\ref{Sec:num}, we study numerically the model introduced in Sec.~\ref{sec:model}, that includes the spin degree of freedom, finite Zeeman field, on-site interactions, and a bound state mediating CAR and ECT between the PMM dots. We still find features that are in qualitative agreement with the predictions from the toy model. However, when adding the spin degree of freedom, the exact spin structure of the low-energy modes (characterized by ``Majorana spin canting angles" in previous works~\cite{Prada_PRB2017,deng2018nonlocality}) plays a role and can influence the appearance of the patterns around the two (different spin-state) crossings. We furthermore study the effect of the interdot Coulomb interaction. Finally, the main conclusions are summarized in Sec.~\ref{Sec:conclusions}.

\section{Model}\label{sec:model}

\begin{figure}[b] \centering
\includegraphics[width=1.0\linewidth]{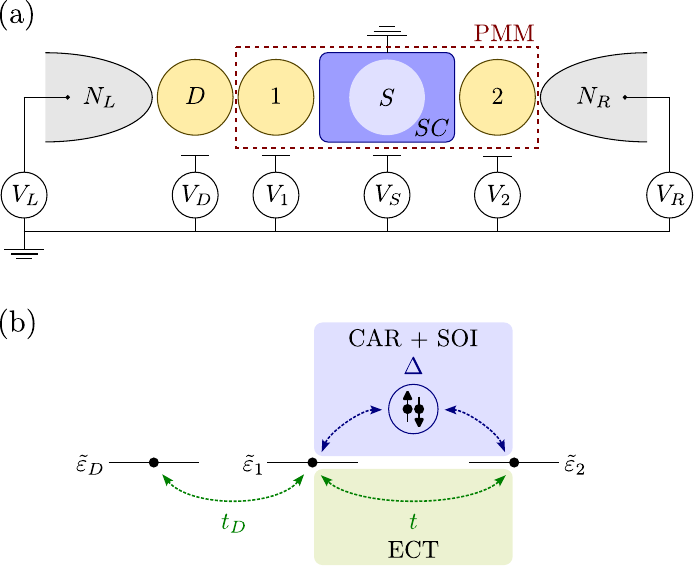}
\caption{(a) Sketch of the dot--PMM system.
Four quantum dots are coupled in series, one of them being strongly proximitized by a grounded superconductor (blue rectangle).
Electrostatic gates control the on-site potentials on all dots and the system is connected to normal-metal reservoirs at its ends, allowing for tunneling spectroscopy.
(b) Cartoon of the simplified model of the four-dot system, where dot $S$ has been integrated out, yielding the effective parameters $t$ and $\Delta$ that capture ECT and CAR processes between dots 1 and 2.
}\label{Fig1}
\end{figure}

The system we consider consists of a linear setup of four tunnel-coupled quantum dots, as sketched in Fig.~\ref{Fig1}(a).
Dot $S$ is proximitized by a grounded superconductor, leading to the appearance of localized Andreev bound states (ABSs). In the regime where the outer PMM dots are weakly coupled to $S$, the ABSs facilitate ECT between them. They also allow for CAR, where a Cooper pair in the superconductor splits up into a correlated singlet pair on dots 1 and 2, or, reversely, a singlet pair on dots 1 and 2 combines into a Cooper pair in the superconductor. \footnote{The notions of ECT and CAR are no longer valid in the regime where the dot $S$ couples strongly to 1 and 2. Nevertheless, the system can then still host PMM sweet spots~\cite{Tsintzis2022}.}
In the presence of a large Zeeman splitting and significant spin-flipping tunneling due to strong spin--orbit interaction, these two processes can effectively mimic the tunneling and $p$-wave pairing terms in a Kitaev chain~\cite{Leijnse_PRB2012,Sau_NatComm2012}.
The part of the system inside the red dashed rectangle in Fig.~\ref{Fig1}(a) can thus behave as a two-site Kitaev chain, a minimal Kitaev chain, with Majorana modes emerging on dots 1 and 2 when the ECT and CAR processes are tuned to be equally strong. The system is coupled to an additional dot, $D$, as indicated in Fig.~\ref{Fig1}, and the combined dot--PMM system is embedded in a transport setup by connecting it to two normal-metal reservoirs $N_{L,R}$.

%%%%%%%%%%%%%%%%%%%%%%%%%%%%%%%%%%%%%%%%%%%%%%%%%%%%%%%%%%%%
%%%%%%%%%%%%%%%%%%%%%%%%%%%%%%%%%%%%%%%%%%%%%%%%%%%%%%%%%%%%

We describe the combined dot--PMM system with the Hamiltonian 
\begin{equation}
    \label{eq:Hsys}
    H_{\rm sys} = H_\mathrm{QDs} + H_\mathrm{ABS} + H_\mathrm{T}\,.
\end{equation}
%$H_{\rm sys} = H_\mathrm{QDs} + H_\mathrm{ABS} + H_\mathrm{T}$.
The first term,
\begin{align}
H_\mathrm{QDs} = {} & {} \sum_{j = D,1,2} \sum_\sigma\varepsilon_{j \sigma} d_{j\sigma}^\dagger d_{j\sigma}
\nonumber\\ {} & {}
+ \sum_{j=D,1,2} \frac{U_j}{2} n_{j}(n_j-1) + V_{D1} n_Dn_1,
\label{eq:QDs}
\end{align}
describes the three normal quantum dots $D$, 1, and 2.
Here, $d_{j\sigma}^\dagger$ is the creation operator for an electron with spin $\sigma=\uparrow,\downarrow$ on dot $j$ and $n_j = d_{j\ua}^\dagger d_{j\ua}+d_{j\da}^\dagger d_{j\da}$ is the number operator on dot $j$.
We assume the orbital level splitting on the dots to be large enough so that it suffices to only consider a single orbital on each dot, meaning that $n_{j} = 0,1,2$.
The on-site potentials $\varepsilon_{j\sigma}$ include a Zeeman splitting $\varepsilon_{j\uparrow,\downarrow} = \varepsilon_j \pm \frac{1}{2}E_{{\rm{Z}}}$, where the $\varepsilon_j$ are assumed to be tunable via the nearby gate electrodes $V_j$, see Fig.~\ref{Fig1}(a).
The on-site charging energy is parametrized by $U_j$ and we also added an interdot charging energy $V_{D1}$ between dots $D$ and 1 (the interaction between dots 1 and 2 is screened by the superconductor).

The virtual coupling between dots 1 and 2 is assumed to be mediated by a discrete ABS on the proximitized dot $S$.
We describe this state with the term
\begin{eqnarray}\label{eq:ABS}
H_\mathrm{ABS} = \sum_{\sigma} \varepsilon_{S\sigma} d_{S\sigma}^\dagger d_{S\sigma} +\Delta_S  d_{S\uparrow}^\dagger d_{S\downarrow}^\dagger + \Delta_S^* d_{S\downarrow} d_{S\uparrow}.
\end{eqnarray}
%where $d_{S\sigma}^\dagger$ is the creation operator for an electron with spin $\sigma$ on dot $S$.
The energies $\varepsilon_{S\uparrow,\downarrow} = \varepsilon_S \pm \frac{1}{2}E_{{\rm{Z}},S}$ include again a Zeeman splitting.
Due to the proximity of the superconductor, (i) the Zeeman energy $E_{{\rm{Z}},S}$ can be renormalized as compared to $E_{\rm Z}$ and (ii) on-site interactions will be efficiently screened, and are therefore neglected.
The superconducting proximity effect is described by the last two terms in Eq.~\eqref{eq:ABS}, where $\Delta_S$ is the induced pairing potential on the dot.

The coupling between neighboring dots is described by
\begin{align}
H_\mathrm{T} =
{} & {}
\sum_{\sigma} \left(t_{D1} d_{D \sigma}^\dagger d_{1{\sigma}}
+ t_{1S} d_{1 \sigma}^\dagger d_{S{\sigma}}
+ t_{S2} d_{S \sigma}^\dagger d_{2{\sigma}} \right) \nonumber\\
{} & {}
+ \sum_{\sigma} s_\sigma\left( t^\mathrm{SO}_{D1} d_{D \sigma}^\dagger d_{1\bar{\sigma}}
+ t^\mathrm{SO}_{1S} d_{1 \sigma}^\dagger d_{S\bar{\sigma}}
+ t^\mathrm{SO}_{S2} d_{S \sigma}^\dagger d_{2\bar{\sigma}}\right) \nonumber\\
{} & {} + \text{H.c.},
\label{eq:HT}
\end{align}
where $\bar\sigma$ is the opposite spin to $\sigma$ and $s_{\ua,\da}=\pm 1$.
The energies $t_{ij}$($t^{\rm SO}_{ij}$) set the amplitude for spin-conserving (spin-flipping) tunneling between dots $i$ and $j$.
In writing $H_{\rm T}$, we have assumed the spin-flipping tunneling to be caused by an effective spin--orbit field $\mathbf{B}_\mathrm{SO}$ that is oriented along the $y$-axis, perpendicular to the direction of the external Zeeman field $\mathbf{B}$, which we have taken to be along $z$, cf.\ Ref.~\cite{Stepanenko2012}; this results in a real Hamiltonian $H_{\rm sys}$.

The full Hamiltonian, including the normal reservoirs, becomes $H = H_{\rm sys} + H_{\rm res} + H_{\rm coup}$, where
\begin{align}
H_{\rm res} = {} & {} \sum_{k, \sigma, r = L,R} \varepsilon_{r k \sigma } c^\dagger_{r k \sigma } c_{r k \sigma },
\end{align}
describes the electrons in the leads, where $\varepsilon_{r k \sigma }$ is the energy of an electron in level $k$ with spin $\sigma$ in lead $r$ (relative to the Fermi level).
The coupling between the system and the reservoirs is described by
\begin{align}
H_{\rm coup} = \sum_{k,\sigma}\left( \lambda_{LD}d^{\dagger}_{D\sigma}c_{L k \sigma } + \lambda_{R2}d^{\dagger}_{2\sigma}c_{R k \sigma } + {\rm H.c.} \right),
\end{align}
where $\lambda_{r j}$ is the coupling strength of the levels in lead $r$ to the level on dot $j$, which we will assume to be spin- and energy-independent for convenience.
This results in a typical tunneling rate to the reservoirs, $\Gamma = 2\pi \lambda^2 \nu_{\rm res}$, with $\nu_{\rm res}$ the effective density of states of the reservoirs.
For our numerical simulations of transport experiments, we always focus on the regime where the tunnelling rates to the reservoirs are the smallest energy scale and solve for the current with a rate-equation approach~\cite{Kirsanskas_CPC2017}. The tunnel couplings between the dots are included in a non-perturbative way and can be strong.

\subsection{Spinless toy model}

A way to simplify the model presented above is to work in the limit of a large Zeeman splitting on all dots and a large induced pairing potential $|\Delta_S|$.
In this case (i) the dots become spin polarized and one can disregard the excited spin state and (ii) the ECT and CAR tunneling processes between dots 1 and 2 can be well described within second-order perturbation theory in $t_{ij}/|\Delta_S|$ and $t_{ij}^{\rm SO}/|\Delta_S|$.
This yields an effectively spinless and non-interacting model, similar to the one used in Ref.~\cite{Leijnse_PRB2012}.

The energy level diagram for the effective model is sketched in Fig.~\ref{Fig1}(b) and described by the simplified Hamiltonian
\begin{align}
\tilde{H}_{\rm sys} = {} & {} \sum_{j=D,1,2} \tilde{\varepsilon}_j d^\dagger_j d_j
+ V_{D1} n_Dn_1
\nonumber\\ {} & {} 
+ \left( t_D d^\dagger_Dd_1 + {t} \, d^\dagger_{1} d_{2} + {\Delta} \, d^\dagger_{1} d^\dagger_{2} + {\rm H.c.}\right).\label{eq:ham_simp}
\end{align}
The parameters $t$ and $\Delta$ are effective, resulting from ECT and CAR processes, respectively, and the on-site energies $\tilde \varepsilon_j$ include a renormalization due to the coupling to the higher-energy states that have been integrated out.

When focusing on the PMM part of the system, i.e., when setting $V_{D1} = t_D = 0$ and disregarding dot $D$, this model reduces to the one used in Ref.~\cite{Leijnse_PRB2012}.
It hosts a so-called sweet spot at $\tilde \varepsilon_1 = \tilde \varepsilon_2 = 0$ and ${\Delta} = {t}$, where the lowest-energy fermionic mode has zero energy and has a corresponding creation operator that can be written as $f^\dagger_0 = \frac{1}{2}\gamma_1 + \frac{i}{2}\tilde\gamma_2$, in terms of the Majorana operators $\gamma_j = d^\dagger_j + d_j$ and $\tilde\gamma_j = i(d^\dagger_j - d_j)$, i.e., the system hosts two well-localized zero-energy Majorana modes.

Away from the sweet spot, the degeneracy of the lowest-energy even and odd modes is lifted.
In the simplified model the energy splitting is
\begin{equation}
    \delta \tilde E_{eo}^{\rm PMM} = \frac{1}{2}\Big[ \sqrt{\tilde \varepsilon_-^2 + 4 t^2} - \sqrt{\tilde \varepsilon_+^2 + 4 \Delta^2} \Big],\label{eq:eo}
\end{equation}
where $\tilde \varepsilon_\pm = \tilde \varepsilon_1 \pm \tilde \varepsilon_2$.
Such a finite energy splitting is relatively straightforward to detect in an experiment, via local tunneling spectroscopy.
Also, the ``Majorana purity'' of the parts of the lowest-energy-mode wave function living on dots 1 and 2 may be reduced when moving away from the sweet spot.
One way to quantify this purity is to introduce the MP on each dot~\cite{Tsintzis2022,tsintzis2023roadmap},
\begin{align}
    \tilde M_j = {} & {} \frac{\langle o | \gamma_j | e \rangle^2 - \langle o | \tilde\gamma_j | e \rangle^2}{\langle o | \gamma_j | e \rangle^2 + \langle o | \tilde\gamma_j | e \rangle^2}
    \nonumber \\
    = {} & {} 
    \frac{-4 t \Delta}{\tilde\varepsilon_+ \tilde \varepsilon_- \mp \sqrt{(\tilde \varepsilon_-^2+4 t^2)(\tilde \varepsilon_+^2 + 4 \Delta^2)}},\label{eq:mp_simp}
\end{align}
where $\ket{o}$ and $\ket{e}$ are the lowest-energy states with even and odd fermion parity, respectively.
Direct signatures of a high MP are much harder to obtain and usually involve identifying fine features in non-local conductance measurements across the PMM system~\cite{Tsintzis2022}.

\section{Results}

\subsection{Analytic results for spinless toy model}\label{sec:anal}

In this work, we explore in detail the possibility to use the extra quantum dot $D$ added to the PMM setup to indirectly assess the Majorana quality of the lowest-energy mode on the PMM system, via local tunneling spectroscopy only.
In the search for Majorana states in proximitized bulk nanowires, the addition of such an extra dot yielded features in the low-energy part of the spectrum that were interpreted as signatures of the absence or presence of low-energy modes with high Majorana localization~\cite{deng2016majorana,Prada_PRB2017,Clarke_PRB2017}.
A numerical investigation using the model introduced in Ref.~\cite{Tsintzis2022} indicated that the dot--PMM setup indeed shows very similar physics to the bulk case, and could potentially reveal information about the quality of the Majorana modes on the PMM part of the system~\cite{tsintzis2023roadmap}.

We thus investigate the full simplified Hamiltonian $\tilde H_{\rm sys}$ as given in Eq.~(\ref{eq:ham_simp}), first setting the interdot interaction to zero, $V_{D1} = 0$, for simplicity.
The two sectors in this Hamiltonian describing the states with total even or odd occupancy of the system are uncoupled. The energy difference $\delta \tilde E_{{\rm{eo}}}$ between the ground states in these two sectors corresponds to the lowest threshold for current to flow through the system~\cite{Clarke_PRB2017}.
At the sweet spot, where $\tilde \varepsilon_1 = \tilde \varepsilon_2 = 0$ and $t=\Delta$, the even and odd Hamiltonians are identical, meaning that $\delta \tilde E_{{\rm{eo}}}=0$, irrespective of the tuning of $\tilde\varepsilon_D$.
In this case one thus expects a robust zero-bias transport signal over an extended range of $\tilde \varepsilon_D$.

When deviating from the sweet spot, either by tuning $\tilde\varepsilon_{1,2}$ away from zero or having $t \neq \Delta$, the even and odd sectors become different.
To capture the leading-order effects of such deviations, we diagonalize the Hamiltonian $\tilde H_{\rm sys}$ analytically at the sweet spot and then treat the deviations perturbatively, calculating the shift of all energy levels up to second order.
This leads to a small splitting between the two ground states,
\begin{align}
\delta \tilde E^{(2)}_{eo} = \tilde\varepsilon_2 \cos \theta+\tau \sin\theta\sin\phi- \frac{\tilde \varepsilon_1\tilde \varepsilon_2}{2t}\sin^3\theta\sin\phi,\label{eq:deltaE_eo}
\end{align}
where we introduced $\tau = t-\Delta$ and the two angles $\phi = \arctan(\tilde\varepsilon_D/t_D)$ and $\theta = \arctan([2t\sqrt{t_D^2+\tilde \varepsilon_D^2}]/t_D^2)$.

%%%%%%%%%%%%%%%%%%%%%%%%%%%%%%%%%%%%%%%%%%%%%%%%%%%%%%%%%%%%
\begin{figure}[t] \centering
\includegraphics[width=1\linewidth]{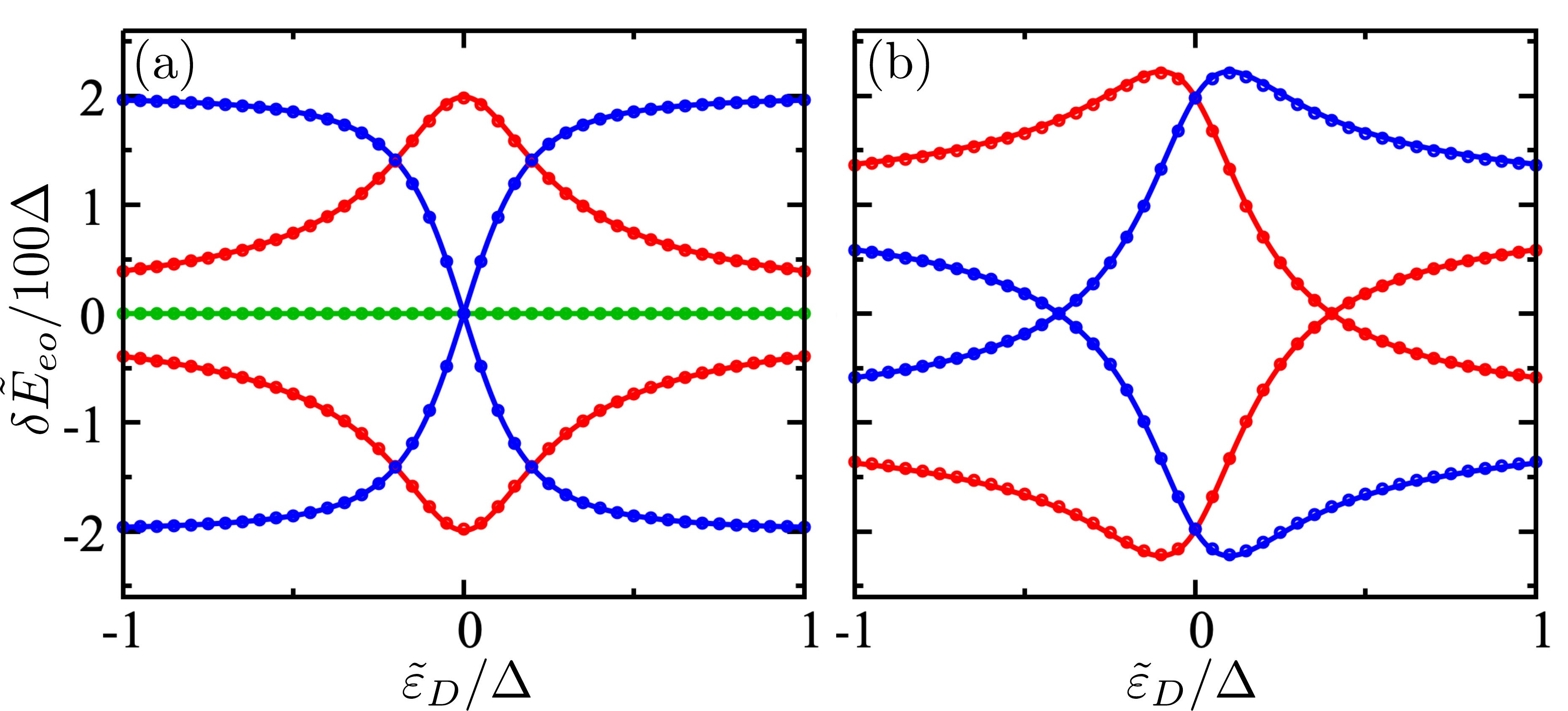}
\caption{Approximate analytical [solid lines, given by Eq.~\eqref{eq:deltaE_eo}] and numerical (dots) calculations of the splitting $\pm \delta \tilde E_{{\rm{eo}}}$ between the even and odd ground states of the Hamiltonian (\ref{eq:ham_simp}) as a function of $\tilde\varepsilon_D$.
(a) The splitting at the sweet spot (green), when $\tilde\varepsilon_2 = 0.2\,\Delta$ (red), and when $\tau =t-\Delta= 0.02\,\Delta$ (blue).
(b) When both on-site potentials $\tilde\varepsilon_{1,2}$ are detuned away from zero the pattern becomes asymmetric; we show $\tilde\varepsilon_1 = 0.1\,\Delta$ and $\tilde\varepsilon_2 = 0.2\,\Delta$ in red and $\tilde\varepsilon_1 = -0.1\,\Delta$ and $\tilde\varepsilon_2 = 0.2\,\Delta$ in blue.
In all plots we have set the tunnel coupling to the extra dot $t_D = 0.2\,\Delta$ and we neglect the interdot charging energy $V_{D1}$.
}\label{FigPCAnal}
\end{figure}
%%%%%%%%%%%%%%%%%%%%%%%%%%%%%%%%%%%%%%%%%%%%%%%%%%%%%%%%%%%%

In Fig.~\ref{FigPCAnal} we show the calculated splitting between the even and odd ground states $\pm \delta \tilde E_{{\rm{eo}}}$ as a function of $\tilde\varepsilon_D$, for several different sets of parameters.
In all plots the solid lines were obtained using the perturbative result of Eq.~(\ref{eq:deltaE_eo}), whereas the dots are the result of numerical diagonalization of the Hamiltonian (\ref{eq:ham_simp}).
Fig.~\ref{FigPCAnal}(a) shows the splitting at the sweet spot (green) and for the case where only one parameter deviates (red: $\tilde\varepsilon_2 = 0.2\,\Delta$, blue: $\tau = 0.02\,\Delta$).
In Fig.~\ref{FigPCAnal}(b) we investigate the splitting when both $\tilde\varepsilon_1$ and $\tilde\varepsilon_2$ are detuned from zero, causing the second-order correction in (\ref{eq:deltaE_eo}) to contribute: in red we show the case $\tilde\varepsilon_1 = 0.1\,\Delta$ and $\tilde\varepsilon_2 = 0.2\,\Delta$ and in blue $\tilde\varepsilon_1 = -0.1\,\Delta$ and $\tilde\varepsilon_2 = 0.2\,\Delta$.
Throughout the plots we used a tunnel coupling $t_D = 0.2\,\Delta$.

First of all, we see that the perturbative expression (\ref{eq:deltaE_eo}) captures the low-energy level structure of the Hamiltonian (\ref{eq:ham_simp}) very well.
Secondly, we note that we observe the same phenomenology in the level structure as for the case of the continuous wire~\cite{Prada_PRB2017,Clarke_PRB2017}, it being flat [green in Fig.~\ref{FigPCAnal}(a)] or showing a ``diamond'' pattern (red in the same Figure), a ``bowtie'' pattern (blue in the same Figure), or something in between [Fig.~\ref{FigPCAnal}(b)], depending on the tuning of $\tilde\varepsilon_{1,2}$, $t$, and $\Delta$.
These different shapes were connected to the interplay between the energy splitting of the two Majorana modes and the degree of their localization on their respective ends of the wire.
A diamond-like line shape indicates a vanishing energy splitting between the Majoranas but a significant direct coupling of the dot level to the Majorana at the far (right) end of the system, whereas a bowtie structure suggests the opposite situation (significant splitting but no coupling of the left dot to the Majorana at the right end of the system)~\cite{Prada_PRB2017}.
Ideal Majorana modes are expected to produce a robust zero-energy response, independent of the tuning of the parameters $\tilde\varepsilon_D$ and $t_D$.

Due to the simplicity of our model, we can make this connection more explicit in our case and test the qualitative statements given above against the analytic expressions for $\delta \tilde E_{eo}^{\rm PMM}$ and $\tilde M_j$ we gave in Eqs.~(\ref{eq:eo},\ref{eq:mp_simp}).
At the sweet spot, the splitting is zero and the MPs are $\pm 1$, as expected.
When $\tilde\varepsilon_2$ is tuned away from zero and $t=\Delta$ [diamond pattern in Fig.~\ref{FigPCAnal}(a)], the splitting of the PMM system remains zero, while $\tilde M_1 = -2t^2/(2t^2 + \tilde\varepsilon_2^2)$ and $\tilde M_2 = 1$. This is consistent with having finite weight of Majorana 2 (the right one) on dot 1, but zero weight of Majorana 1 on dot 2. In this case, the extra dot can couple to both Majorana modes and thus hybridize them, lifting the ground state degeneracy close to $\tilde{\varepsilon}_D=0$. In contrast, when instead $\tilde\varepsilon_1$ is tuned away from zero the even-odd degeneracy is not affected, as the extra dot now couples to the dot with unit MP, meaning that it couples only to one of the Majorana modes.
When both on-site energies are still tuned to zero but there is a finite mismatch $\tau = t-\Delta \neq 0$ [bowtie shape in Fig.~\ref{FigPCAnal}(a)], then we find $\tilde M_{1,2} = \mp 1$ but now $\delta \tilde E^{\rm PMM}_{eo} = \tau\neq0$. Therefore, the Majoranas are well-localized in this case, although they have a finite energy splitting. This results in a constant even--odd energy splitting, except near $\tilde{\varepsilon}_D=0$.

We thus corroborate the findings and their interpretation presented in Refs.~\cite{Prada_PRB2017,Clarke_PRB2017}, adding deeper insight in the underlying physics through our simple analytic results.
Furthermore, this suggests that, analogously to the case of the continuous wire, investigating the low-energy level structure of a combined dot--PMM system through straightforward local tunneling spectroscopy can in principle allow to independently determine the residual splitting between the PMM Majorana modes, as well as their degree of localization on the two ends of the system.

However, if no special care is taken with the design of the experiment, $V_{D1}$ will in general be non-zero, since there are no superconducting elements in between dots $D$ and 1 that would screen the interactions.
The nearest-neighbor interdot charging energy is of the order of $\sim e^2/(4\pi \varepsilon_r \varepsilon_0 d)$ with $\varepsilon_r\varepsilon_0$ the dielectric permittivity of the surroundings and $d$ the interdot distance, yielding typically $V_{D1}\sim 0.1$--1~meV, which can make $V_{D1}$ in fact a dominating energy scale in our model.
To understand the effect of significant interdot interactions, we thus revisit our simplified Hamiltonian (\ref{eq:ham_simp}) now assuming $V_{D1}$ to be large.
In that limit, the two states $\ket{1;10}$ and $\ket{1;11}$ (using the notation $\ket{n_D;n_1n_2}$) will have a much higher energy than the other six states we consider and can thus be disregarded.

%%%%%%%%%%%%%%%%%%%%%%%%%%%%%%%%%%%%%%%%%%%%%%%%%%%%%%%%%%%%
\begin{figure}[t] \centering
\includegraphics[width=1\linewidth]{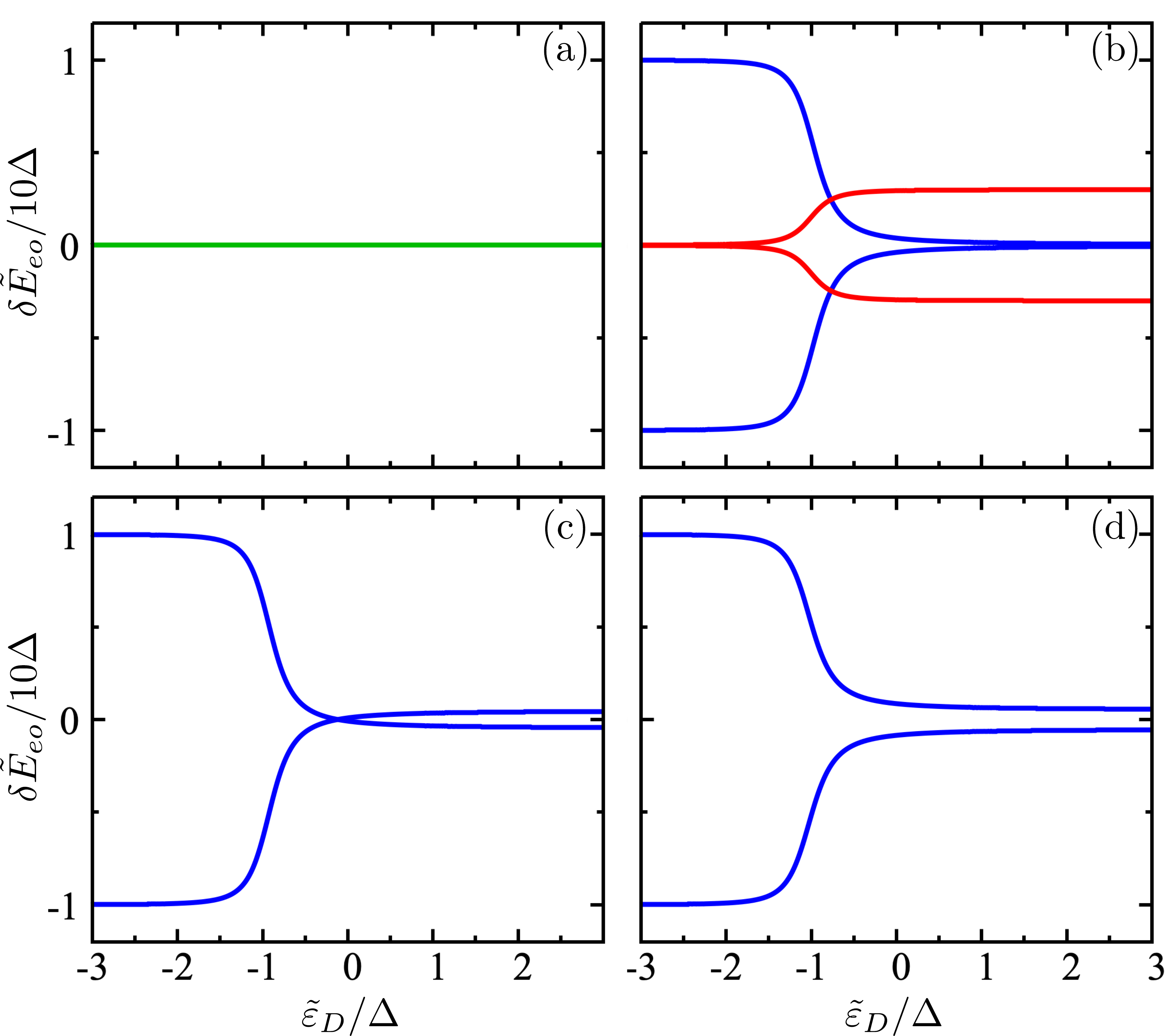}
\caption{Energy splitting $\delta \tilde E_{eo}$ between the lowest even and odd state of $\tilde H_{\rm sys}$ as given by (\ref{eq:ham_simp}) in the limit of large $V_{D1}$.
(a) At the sweet spot $\tilde\varepsilon_1 = \tilde \varepsilon_2 = \tau = 0$ we still see a persistent degeneracy.
(b) Tuning $\tilde\varepsilon_2 = 0.1\,\Delta$ (blue line) or $\tau = 0.03\,\Delta$ (red line) away from the sweet spot no longer results in the characteristic diamond or bowtie patterns, but rather a degeneracy on one side of the plot that is split up on the other side.
Tuning both $\tilde\varepsilon_{1}$ and $\tilde\varepsilon_{2}$ away from zero [$\tilde\varepsilon_1 = -\tilde\varepsilon_2 = 0.1\,\Delta$ in (c); $\tilde\varepsilon_1 = \tilde\varepsilon_2 = 0.1\,\Delta$ in (d)] results in a finite splitting both at large positive and negative $\tilde \varepsilon_D$, with a zero crossing around $\tilde \varepsilon_D \approx 0$ only for different signs of $\tilde\varepsilon_1$ and $\tilde \varepsilon_2$.
In all plots we have set $t_D = 0.2\,\Delta$.
}\label{FigPCInt}
\end{figure}
%%%%%%%%%%%%%%%%%%%%%%%%%%%%%%%%%%%%%%%%%%%%%%%%%%%%%%%%%%%%

The remaining three-level Hamiltonians for the even and odd sectors can straightforwardly be diagonalized and in Fig.~\ref{FigPCInt} we show the calculated splitting between the even and odd ground states for different sets of parameters, all close to the sweet spot.
Fig.~\ref{FigPCInt}(a) confirms that the splitting consistently vanishes at the sweet spot, independent of $\tilde\varepsilon_D$.
Indeed, the effect of the interaction can be understood as a conditional upward shift of the level $\tilde\varepsilon_1$ depending on the occupation of dot $D$, and we thus do not expect a finite $\delta \tilde E_{{\rm{eo}}}$ to emerge at the sweet spot as long as $\tilde\varepsilon_2 = 0$ [see Eq.~({\ref{eq:deltaE_eo})].
When we tune the parameters away from the sweet spot [Fig.~\ref{FigPCInt}(b--d), see the caption for plot parameters] asymmetric patterns of $\delta \tilde E_{{\rm{eo}}}$ appear.

The structure of these patterns can easily be understood by investigating the $\tilde\varepsilon_D$-dependent level structure in more detail.
At large negative $\tilde\varepsilon_D$ the even and odd ground states are the states that involve occupation of the extra dot, $\ket{1;01}$ and $\ket{1;00}$, respectively.
The even--odd ground state splitting is in this case always equal to $\tilde\varepsilon_2$.
In the limit of large and positive $\tilde\varepsilon_D$, the states $\ket{1;01}$ and $\ket{1;00}$ now have high energy and can thus be disregarded.
What is left are the states $\ket{0;00}$, $\ket{0;11}$, $\ket{0;10}$, and $\ket{0;01}$, i.e., the system is equivalent to the original PMM system without the extra dot attached; the ground state splitting for large positive $\tilde\varepsilon_D$ then simply becomes $\delta \tilde E_{{\rm{eo}}}^{\rm PMM}$ as given by Eq.~(\ref{eq:eo}).
We thus find the limiting values for $\delta \tilde E_{{\rm{eo}}}$
\begin{align}
    \delta \tilde E_{{\rm{eo}}} = \begin{cases}
    \tilde\varepsilon_2 & \text{ for } \tilde\varepsilon_D \to -\infty, \\
    \delta \tilde E_{{\rm{eo}}}^{\rm PMM} & \text{ for } \tilde\varepsilon_D \to \infty.
    \end{cases}\label{eq:limed}
\end{align}
The effect from the non-local interaction between the dots $D$ and $1$ when $D$ is filled can be compensated by shifting the energy of dot $1$.

Comparing this result to Fig.~\ref{FigPCInt}, we see that it explains the limiting values of all patterns of $\delta \tilde E_{eo}$, both at the sweet spot and away from it.
The subtle difference between Figs.~\ref{FigPCInt}(c) and (d) (the presence or absence of an apparent zero crossing of $\delta \tilde E_{eo}$) can also be understood from Eq.~(\ref{eq:limed}), by comparing the sign of the two limiting values.
Indeed, when $\tilde\varepsilon_1 = \tilde\varepsilon_2 = 0.1\,\Delta$ and $t=\Delta$ [Fig.~\ref{FigPCInt}(c)] we have $\tilde\varepsilon_2 > 0$ and $\delta \tilde E_{eo}^{\rm PMM} <0$ [see Eq.~(\ref{eq:eo})], whereas for $\tilde\varepsilon_1 = -\tilde\varepsilon_2 = -0.1\,\Delta$ and $t=\Delta$ [Fig.~\ref{FigPCInt}(c)] one finds that both $\tilde\varepsilon_2, \delta \tilde E_{eo}^{\rm PMM} > 0$.

These results thus suggest that within the simplified model the sweet spot can still be distinguished in the presence of significant interdot Coulomb interactions, since the sweet spot is the only point in parameter space where $\delta \tilde E_{eo}$ remains zero regardless of the value of $\tilde \varepsilon_D$.

\subsection{Numerical results}
\label{Sec:num}

We now turn to the full model presented in Sec.~\ref{sec:model}, to investigate how much of the phenomenology described above survives in a spinful model with finite Zeeman energies and on-site Coulomb interactions. To understand the transport through the system, we diagonalize $H_{\rm sys}$~\eqref{eq:Hsys}, sketched in Fig.~\ref{Fig1}, including finite charging energy, and use rate equations to calculate transport.
In this case, to assess the Majorana quality of the eigenmodes, we need to use a more general version of a MP that includes both spin states.
We will thus use the MP defined as~\cite{Sedlmayr2015,Sedlmayr2016,Aksenov2020,Tsintzis2022}
\begin{equation}
M_j = \frac{ \sum_\sigma \left( w_\sigma^2 - z_\sigma^2 \right)}{\sum_\sigma \left( w_\sigma^2 + z_\sigma^2 \right)}, \label{eq:mp}\\
\end{equation}
with $w_\sigma = \langle o | (d_{j\sigma} + d_{j\sigma}^\dagger) |e\rangle$ and $z_\sigma = \langle o | (d_{j\sigma} - d_{j\sigma}^\dagger) |e\rangle$, to quantify the Majorana quality of the lowest-energy mode. 
In the presence of interactions and finite Zeeman energy, true sweet spots with both an even--odd ground-state degeneracy and ideal Majoranas in each of the dots, {i.e.}, $|M_{1,2}|=1$, do not exist. However, it is possible to find sweet spots with even--odd degeneracy and MP well over 0.95 for experimentally relevant parameters~\cite{Tsintzis2022}. 

\begin{figure}[t] \centering
\includegraphics[width=1\linewidth]{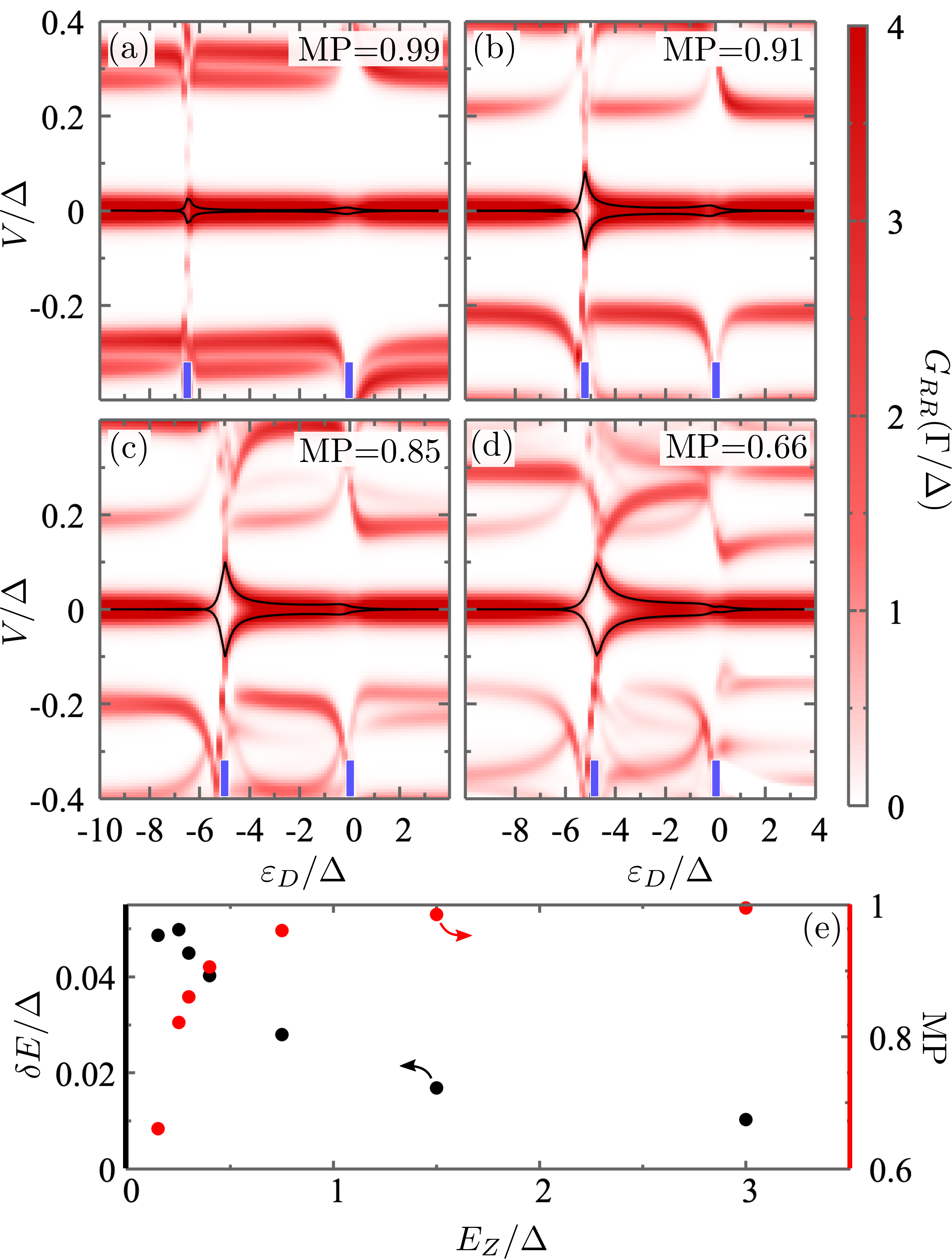}
\caption{(a--d) Numerically calculated local conductance measured at the right end of the system, for different MP. The black lines show the voltage threshold for zero temperature, i.e., $|V|=\delta E_{eo}$. The blue tics mark the energies where a spin-split level on the additional dot crosses zero energy. (e) Maximum energy splitting (black dots) and MP (red) for increasing Zeeman splitting. In all panels, we used $U_D=U_1=U_2=5\,\Delta$, $U_S=V_{D1}=E_{{\rm Z},S}=0$, $t_{1S}=t_{2S}=2t_{D1}=0.5\,\Delta$, and $t^{\rm SO}_{1S}=t^{\rm SO}_{2S}=2t^{\rm SO}_{D1}=0.1\,\Delta$. We vary the Zeeman splitting between the different panels, choosing $E_{\rm Z}=1.5$, 0.4, 0.25, and 0.15 (all in units of $\Delta$) for panels (a--d). The configurations of all gate-tuning parameters are given in Table \ref{table1} in App.~\ref{app:parameters} for every sweet spot. To calculate the conductance, we assume the tunnel rates to the leads, $\Gamma$, to be the smallest energy scale and we set the temperature to $T=0.005\,\Delta$.
The results shown in (a,d,e) were also presented in Ref.~\cite{tsintzis2023roadmap}.
}\label{Fig:MP}
\end{figure}

First, we will investigate how the additional dot can be used to distinguish these high-MP sweet spots from situations with lower MP.
In the ideal case, the outer dots of the PMM system host perfectly localized Majorana states, and sweeping the additional level $\varepsilon_D$ cannot split the even--odd degeneracy, since dot $D$ couples to only one of the two states~\cite{Prada_PRB2017,Clarke_PRB2017}. In the case of high-quality sweet spots with near-unit polarization, we still expect the ground-state splitting to be strongly reduced and indiscernible in transport measurements, due to, e.g., finite resolution and temperature.

In Fig.~\ref{Fig:MP} we show numerical results exploring the maximum splitting observed as a function of MP.
In all cases, we tuned the PMM system to a sweet spot with even--odd degeneracy and maximal MP, varying the Zeeman splitting across Fig.~\ref{Fig:MP}(a--d).
We plot the calculated conductance measured at the right side of the system as a function of $\varepsilon_D$ and applied bias voltage $V$, for decreasing $E_{\rm Z}$.
We see that decreasing the Zeeman splitting (i) brings down in energy excited spin states, (ii) lowers the quality of the sweet spot, see Figs.~\ref{Fig:MP}(e), and (iii) increases the maximum ground-state splitting observed in the conductance.
Indeed, when the MP is substantially lower than 1, the lowest fermionic mode on the PMM system no longer separates into well-localized Majorana states on the outer dots.
%Instead, states appear in the outer PMM QDs composed by an even number of local Majorana states with zero overlap integral.
In this case, the connection to an external quantum system can split the even--odd degeneracy by coupling to both Majorana components of the mode, thereby effectively yielding a coupling between the two Majoranas and thus lifting their degeneracy.
We note that this is generally true, independent of the nature of the additional quantum system, as long as it can mediate a coherent coupling between the Majoranas.
The lowest conductance line shows a diamond-like pattern with a maximum when one of the spin-split levels of the additional dot crosses zero energy.
Decreasing the MP increases the maximum splitting in the conductance, illustrated in Fig.~\ref{Fig:MP}(e), where we show the maximum ground state splitting (black dots) and MP (red dots) as a function of the Zeeman splitting.
These results confirm that, also in the more realistic case, an insensitivity of the ground-state degeneracy to $\varepsilon_D$ indicates a high MP on the PMM system.

Previously, the non-local conductance has served as a main tool for identifying Majorana sweet spots in minimal Kitaev chains~\cite{Dvir2023}. However, discerning between high- and low-MP sweet spots has proven challenging due to their similarities in non-local conductance patterns. The results presented in Fig.~\ref{Fig:MP} show that the additional dot we introduced to the setup offers complementary insights into the localization of the Majorana state, resulting in distinct qualitative outcomes for high- and low-MP sweet spots, also for the more realistic case with finite Zeeman and Coulomb energy. 

\begin{figure}[t] \centering
\includegraphics[width=1\linewidth]{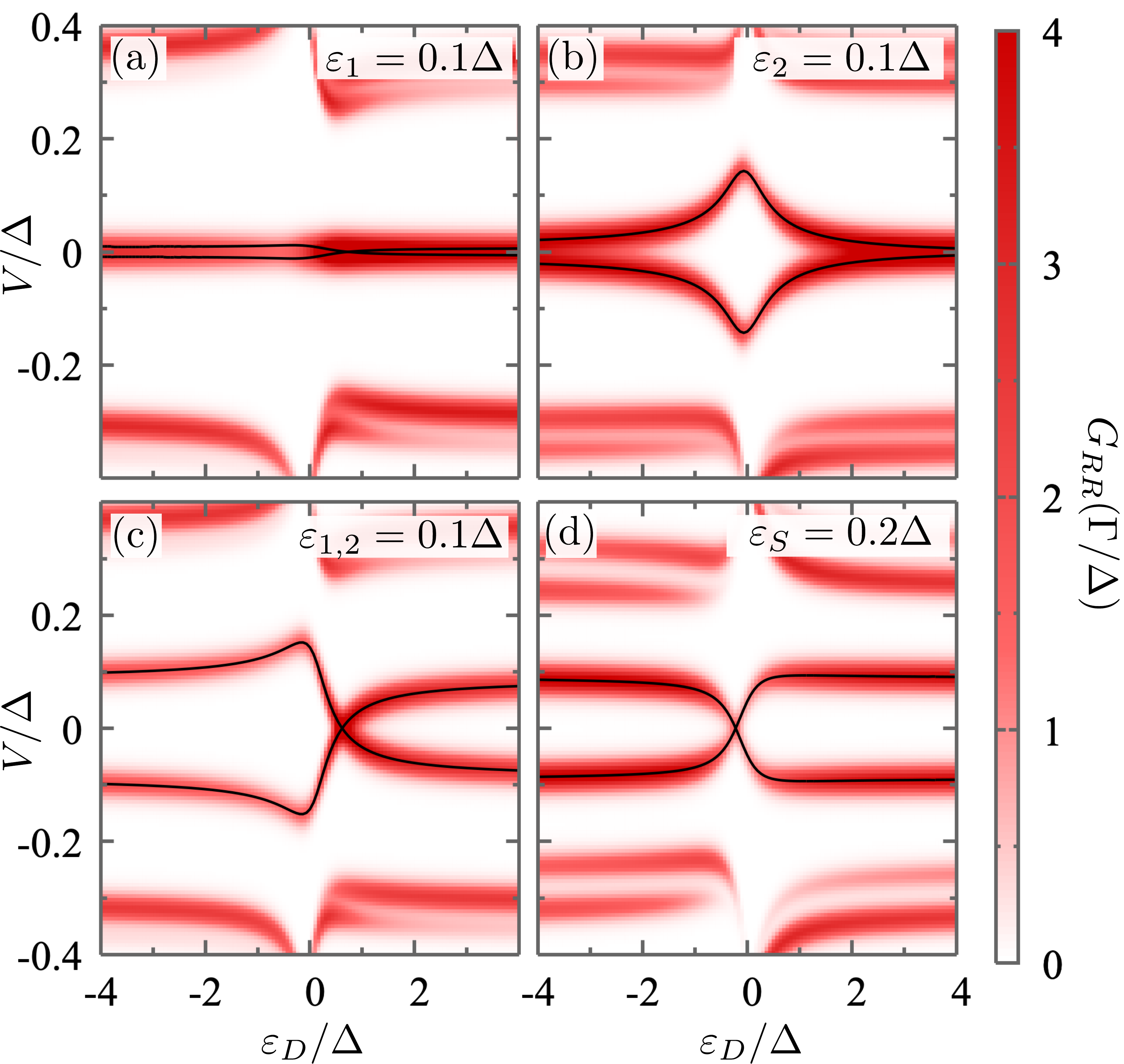}
\caption{Conductance through the right dot for a high-MP PMM. Parameters are the same as in Fig.~\ref{Fig:MP}(a), except for the variables indicated in the top right of each panel, that are shifted away from the sweet spot the by the value given.
}\label{Fig:SP_detuning}
\end{figure}

We now turn to the behavior of the conductance spectrum away from the sweet spot, comparing it to the simple analytic results presented in Sec.~\ref{sec:anal}, see also the blue and red lines in Fig.~\ref{FigPCAnal}.
%PMMs offer a way to control the Majorana weight in both dots, providing a way to increase confidence on the Majorana character of the found sweet spot. In a good sweet spot, two non-overlapping Majorna states localize at the two end of the PMM system. Therefore, an additional QD attached to the system does not split the energy degeneracy. 
Detuning the energy level of one of the dots away from the sweet spot increases the weight of the ``detuned'' Majorana on the opposite dot without lifting the ground state degeneracy~\cite{Leijnse_PRB2012}.
The additional dot, sensitive to the local Majorana components, can lift the even--odd degeneracy by providing an effective coupling between the two Majoranas.
This is confirmed in Figs.~\ref{Fig:SP_detuning}(a,b), showing that the extra quantum dot coupled to the left side of the PMM system is only sensitive to deviations in the right dot of the PMM system.
The splitting of the ground state follows the same diamond-like pattern as found from the toy model, see Fig.~\ref{FigPCAnal}(a) and is qualitatively similar to the pattern found for low-MP sweet spots, see Fig.~\ref{Fig:MP}.
Shifting the energy of both dots of the PMM system lifts the even--odd degeneracy, leading to a bowtie-like pattern in the local conductance, as illustrated in Fig.~\ref{Fig:SP_detuning}(c), cf.~Fig.~\ref{FigPCAnal}(b), where the relative splitting at negative and positive $\varepsilon_D$ depends again on the sign of $\varepsilon_{1,2}$. 
A similar pattern is observed when the relative values of the coupling in the even and odd subspaces (CAR and ECT in the weak tunneling regime) are changed, see Fig.~\ref{Fig:SP_detuning}(d), achieved by detuning the energy level of the central PMM dot $\varepsilon_S$, that controls the properties of the central ABS.
Again, the pattern looks very similar to the one predicted by the toy model, see the blue results in Fig.~\ref{FigPCAnal}(a).

\begin{figure}[t] \centering
\includegraphics[width=1\linewidth]{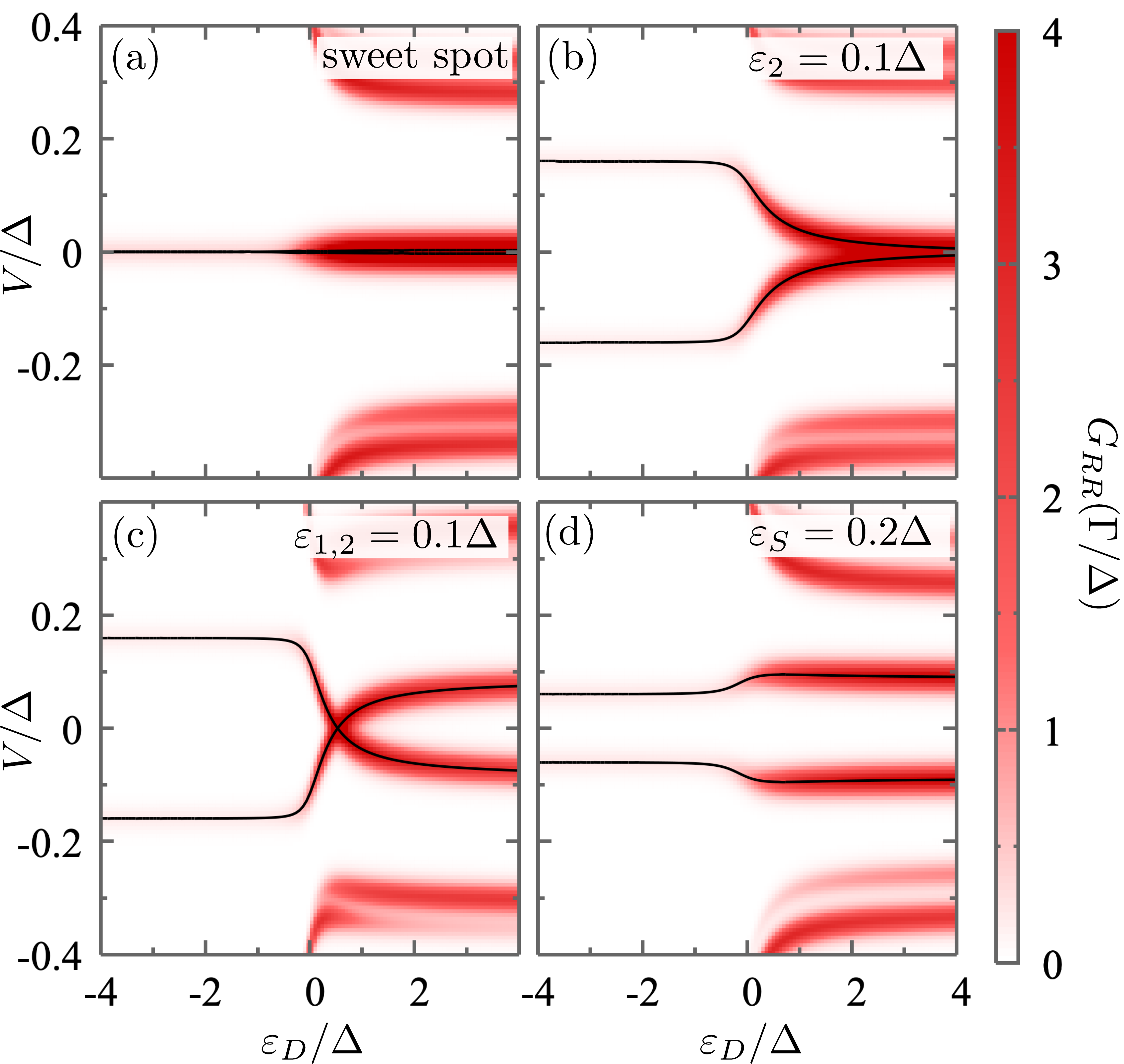}
\caption{Conductance through the right dot for a high-MP PMM. Parameters are the same as in Fig.~\ref{Fig:MP}(a), with $V_{D1}=0.4\,U_2$, except for the parameters indicated in the top right of each panel.
}\label{Fig:V_LD}
\end{figure}

We finally explore the role of interdot Coulomb interactions, again assuming that the interactions between dots 1 and 2 are strongly screened by the grounded superconductor and thus only considering interactions between dots $D$ and $1$, see Fig.~\ref{Fig1}(a).
The results are shown in Fig.~\ref{Fig:V_LD} and are again qualitatively similar to the results found for the toy model investigated in Sec.~\ref{sec:anal}.
At a high-MP sweet spot, the additional Coulomb interaction cannot split the ground-state degeneracy, see Fig.~\ref{Fig:V_LD}(a); it merely leads to a shift of the energy level on dot 1, depending on the occupation of dot $D$, which does not affect the degeneracy, as explained above.
It does, however, quench the local conductance when the additional dot is filled, as can be clearly seen from the reduced intensity of the conductance peaks at negative $\varepsilon_D$. This is due to an effective energy renormalization of the left PMM dot that brings it away from the sweet spot.
Away from the sweet spot [Figs.~\ref{Fig:V_LD}(b--d)] the non-local charging energy affects the conductance patterns in a very similar way to what we found from the toy model in Sec.~\ref{sec:anal}.
For instance, the diamond-like pattern, originating from two Majorana states having finite weight in the same dot due to either a non-ideal MP, opens up for negative $\varepsilon_D$, as shown in Fig.~\ref{Fig:V_LD}(b).
We again explain this behavior in terms of an effective detuning of $\varepsilon_1$ due to the interaction with the electron on dot $D$.
Similarly, the bowtie pattern found for $\varepsilon_1=\varepsilon_2$ becomes asymmetric in $\varepsilon_D$, due to a renormalization of the dot energy when the additional dot $D$ fills up, see Fig.~\ref{Fig:V_LD}(c).
Finally, the bowtie pattern can also open up when the CAR and ECT amplitudes become different, which we do by detuning $\varepsilon_S$, see Fig.~\ref{Fig:V_LD}(d).
In all of the described cases, the phenomenology can again be understood in terms of the simple analytic model investigated in Sec.~\ref{sec:anal}, and the limiting splittings we observe for large $|\varepsilon_D|$ are consistent with Eq.~\eqref{eq:limed}.

We thus conclude from our numerical calculations that the analytic insight in the structure of the even--odd splitting we gained in Sec.~\ref{sec:anal} still provides a qualitative understanding of the underlying physics of the dot--PMM system when including realistic ingredients such as finite Zeeman splitting, strong coupling between the PMM dots, and Coulomb interactions.

\section{Conclusion}
\label{Sec:conclusions}

In this work, we have shown that an additional quantum dot can help assessing the localization of Majorana states in minimal Kitaev chains. We have shown that adding an additional quantum dot to the chain can help distinguishing situations with well-localized Majoranas from other cases where an even number of Majoranas overlap. In the latter case, the additional dot couples to the overlapping Majorana states, thereby lifting the even--odd degeneracy, which is measurable using local transport or microwave measurements. We have performed analytic and numerical calculations to understand the change of the ground state due to the coupling to the additional dot and we studied its effect on the conductance. The system remains degenerate for the case where the dot couples to a single Majorana state, resulting in a robust zero-energy feature, and the degeneracy splits when the dot couples to multiple Majorana states, showing either a diamond or a bowtie shape, depending on whether the PMM system's ground state is degenerate when the measurement dot is decoupled.

This work contributes to the efforts on finding and characterizing Majorana sweet spots in minimal Kitaev chains, and provides a deeper understanding of the physics underlying the coupling of external quantum dots to Majorana states. Distinguishing between true sweet spots and other energy crossings where the local wavefunction does not have a Majorana character is a central question in the field, and will be essential for progress toward experiments demonstrating the presence of non-abelian quasiparticles in condensed matter systems.

\section{Acknowledgments}

We thank Karsten Flensberg for very helpful discussions, and Ramón Aguado and D. Míchel Pino for valuable feedback on the manuscript.
This work has received funding from the European Research Council (ERC) under the European Unions Horizon 2020 research and innovation programme under Grant Agreement No. 856526, the Spanish CM “Talento Program” (project No. 2022-T1/IND-24070), the Swedish Research Council under Grant Agreement No. 2020-03412, the European Union’s Horizon 2020 research and innovation program under the Marie Sklodowska-Curie Grant Agreement No. 10103324, and NanoLund.

\appendix

\bibliographystyle{apsrev4-2}
\bibliography{bibliography}

\begin{thebibliography}{57}%
\makeatletter
\providecommand \@ifxundefined [1]{%
 \@ifx{#1\undefined}
}%
\providecommand \@ifnum [1]{%
 \ifnum #1\expandafter \@firstoftwo
 \else \expandafter \@secondoftwo
 \fi
}%
\providecommand \@ifx [1]{%
 \ifx #1\expandafter \@firstoftwo
 \else \expandafter \@secondoftwo
 \fi
}%
\providecommand \natexlab [1]{#1}%
\providecommand \enquote  [1]{``#1''}%
\providecommand \bibnamefont  [1]{#1}%
\providecommand \bibfnamefont [1]{#1}%
\providecommand \citenamefont [1]{#1}%
\providecommand \href@noop [0]{\@secondoftwo}%
\providecommand \href [0]{\begingroup \@sanitize@url \@href}%
\providecommand \@href[1]{\@@startlink{#1}\@@href}%
\providecommand \@@href[1]{\endgroup#1\@@endlink}%
\providecommand \@sanitize@url [0]{\catcode `\\12\catcode `\$12\catcode
  `\&12\catcode `\#12\catcode `\^12\catcode `\_12\catcode `\%12\relax}%
\providecommand \@@startlink[1]{}%
\providecommand \@@endlink[0]{}%
\providecommand \url  [0]{\begingroup\@sanitize@url \@url }%
\providecommand \@url [1]{\endgroup\@href {#1}{\urlprefix }}%
\providecommand \urlprefix  [0]{URL }%
\providecommand \Eprint [0]{\href }%
\providecommand \doibase [0]{https://doi.org/}%
\providecommand \selectlanguage [0]{\@gobble}%
\providecommand \bibinfo  [0]{\@secondoftwo}%
\providecommand \bibfield  [0]{\@secondoftwo}%
\providecommand \translation [1]{[#1]}%
\providecommand \BibitemOpen [0]{}%
\providecommand \bibitemStop [0]{}%
\providecommand \bibitemNoStop [0]{.\EOS\space}%
\providecommand \EOS [0]{\spacefactor3000\relax}%
\providecommand \BibitemShut  [1]{\csname bibitem#1\endcsname}%
\let\auto@bib@innerbib\@empty
%</preamble>
\bibitem [{\citenamefont {Kitaev}(2001)}]{Kitaev_2001}%
  \BibitemOpen
  \bibfield  {author} {\bibinfo {author} {\bibfnamefont {A.~Y.}\ \bibnamefont
  {Kitaev}},\ }\href {https://doi.org/10.1070/1063-7869/44/10s/s29} {\bibfield
  {journal} {\bibinfo  {journal} {Physics-Uspekhi}\ }\textbf {\bibinfo {volume}
  {44}},\ \bibinfo {pages} {131} (\bibinfo {year} {2001})}\BibitemShut
  {NoStop}%
\bibitem [{\citenamefont {Leijnse}\ and\ \citenamefont
  {Flensberg}(2012{\natexlab{a}})}]{Leijnse_Review2012}%
  \BibitemOpen
  \bibfield  {author} {\bibinfo {author} {\bibfnamefont {M.}~\bibnamefont
  {Leijnse}}\ and\ \bibinfo {author} {\bibfnamefont {K.}~\bibnamefont
  {Flensberg}},\ }\href {https://doi.org/10.1088/0268-1242/27/12/124003}
  {\bibfield  {journal} {\bibinfo  {journal} {Semicond. Sci. Technol.}\
  }\textbf {\bibinfo {volume} {27}},\ \bibinfo {pages} {124003} (\bibinfo
  {year} {2012}{\natexlab{a}})}\BibitemShut {NoStop}%
\bibitem [{\citenamefont {Alicea}(2012)}]{Alicea_RPP2012}%
  \BibitemOpen
  \bibfield  {author} {\bibinfo {author} {\bibfnamefont {J.}~\bibnamefont
  {Alicea}},\ }\href {https://doi.org/10.1088/0034-4885/75/7/076501} {\bibfield
   {journal} {\bibinfo  {journal} {Rep. Prog. Phys.}\ }\textbf {\bibinfo
  {volume} {75}},\ \bibinfo {pages} {076501} (\bibinfo {year}
  {2012})}\BibitemShut {NoStop}%
\bibitem [{\citenamefont {Beenakker}(2013)}]{beenakker2013search}%
  \BibitemOpen
  \bibfield  {author} {\bibinfo {author} {\bibfnamefont {C.~W.~J.}\
  \bibnamefont {Beenakker}},\ }\href
  {https://doi.org/10.1146/annurev-conmatphys-030212-184337} {\bibfield
  {journal} {\bibinfo  {journal} {Annu. Rev. Condens. Matter Phys.}\ }\textbf
  {\bibinfo {volume} {4}},\ \bibinfo {pages} {113} (\bibinfo {year}
  {2013})}\BibitemShut {NoStop}%
\bibitem [{\citenamefont {Aguado}(2017)}]{Aguado_Nuovo2017}%
  \BibitemOpen
  \bibfield  {author} {\bibinfo {author} {\bibfnamefont {R.}~\bibnamefont
  {Aguado}},\ }\href {https://doi.org/10.1393/ncr/i2017-10141-9} {\bibfield
  {journal} {\bibinfo  {journal} {Riv. Nuovo Cimento}\ }\textbf {\bibinfo
  {volume} {40}},\ \bibinfo {pages} {523} (\bibinfo {year} {2017})}\BibitemShut
  {NoStop}%
\bibitem [{\citenamefont {Beenakker}(2020)}]{BeenakkerReview_20}%
  \BibitemOpen
  \bibfield  {author} {\bibinfo {author} {\bibfnamefont {C.~W.~J.}\
  \bibnamefont {Beenakker}},\ }\href
  {https://scipost.org/10.21468/SciPostPhysLectNotes.15} {\bibfield  {journal}
  {\bibinfo  {journal} {SciPost Phys. Lect. Notes 15}\ } (\bibinfo {year}
  {2020})}\BibitemShut {NoStop}%
\bibitem [{\citenamefont {Flensberg}\ \emph {et~al.}(2021)\citenamefont
  {Flensberg}, \citenamefont {von Oppen},\ and\ \citenamefont
  {Stern}}]{flensberg2021engineered}%
  \BibitemOpen
  \bibfield  {author} {\bibinfo {author} {\bibfnamefont {K.}~\bibnamefont
  {Flensberg}}, \bibinfo {author} {\bibfnamefont {F.}~\bibnamefont {von
  Oppen}},\ and\ \bibinfo {author} {\bibfnamefont {A.}~\bibnamefont {Stern}},\
  }\href {https://doi.org/https://doi.org/10.1038/s41578-021-00336-6}
  {\bibfield  {journal} {\bibinfo  {journal} {Nat. Rev. Mat.}\ }\textbf
  {\bibinfo {volume} {6}},\ \bibinfo {pages} {944} (\bibinfo {year}
  {2021})}\BibitemShut {NoStop}%
\bibitem [{\citenamefont {Marra}(2022)}]{Marra_Review2022}%
  \BibitemOpen
  \bibfield  {author} {\bibinfo {author} {\bibfnamefont {P.}~\bibnamefont
  {Marra}},\ }\href {https://doi.org/10.1063/5.0102999} {\bibfield  {journal}
  {\bibinfo  {journal} {J. Appl. Phys.}\ }\textbf {\bibinfo {volume} {132}},\
  \bibinfo {pages} {231101} (\bibinfo {year} {2022})}\BibitemShut {NoStop}%
\bibitem [{\citenamefont {Oreg}\ \emph {et~al.}(2010)\citenamefont {Oreg},
  \citenamefont {Refael},\ and\ \citenamefont {von Oppen}}]{Oreg_PRL2010}%
  \BibitemOpen
  \bibfield  {author} {\bibinfo {author} {\bibfnamefont {Y.}~\bibnamefont
  {Oreg}}, \bibinfo {author} {\bibfnamefont {G.}~\bibnamefont {Refael}},\ and\
  \bibinfo {author} {\bibfnamefont {F.}~\bibnamefont {von Oppen}},\ }\href
  {https://doi.org/10.1103/PhysRevLett.105.177002} {\bibfield  {journal}
  {\bibinfo  {journal} {Phys. Rev. Lett.}\ }\textbf {\bibinfo {volume} {105}},\
  \bibinfo {pages} {177002} (\bibinfo {year} {2010})}\BibitemShut {NoStop}%
\bibitem [{\citenamefont {Lutchyn}\ \emph {et~al.}(2010)\citenamefont
  {Lutchyn}, \citenamefont {Sau},\ and\ \citenamefont
  {Das~Sarma}}]{Lutchyn_PRL2010}%
  \BibitemOpen
  \bibfield  {author} {\bibinfo {author} {\bibfnamefont {R.~M.}\ \bibnamefont
  {Lutchyn}}, \bibinfo {author} {\bibfnamefont {J.~D.}\ \bibnamefont {Sau}},\
  and\ \bibinfo {author} {\bibfnamefont {S.}~\bibnamefont {Das~Sarma}},\ }\href
  {https://doi.org/10.1103/PhysRevLett.105.077001} {\bibfield  {journal}
  {\bibinfo  {journal} {Phys. Rev. Lett.}\ }\textbf {\bibinfo {volume} {105}},\
  \bibinfo {pages} {077001} (\bibinfo {year} {2010})}\BibitemShut {NoStop}%
\bibitem [{\citenamefont {Lutchyn}\ \emph {et~al.}(2018)\citenamefont
  {Lutchyn}, \citenamefont {Bakkers}, \citenamefont {Kouwenhoven},
  \citenamefont {Krogstrup}, \citenamefont {Marcus},\ and\ \citenamefont
  {Oreg}}]{Lutchyn_NatRev2018}%
  \BibitemOpen
  \bibfield  {author} {\bibinfo {author} {\bibfnamefont {R.~M.}\ \bibnamefont
  {Lutchyn}}, \bibinfo {author} {\bibfnamefont {E.~P. A.~M.}\ \bibnamefont
  {Bakkers}}, \bibinfo {author} {\bibfnamefont {L.~P.}\ \bibnamefont
  {Kouwenhoven}}, \bibinfo {author} {\bibfnamefont {P.}~\bibnamefont
  {Krogstrup}}, \bibinfo {author} {\bibfnamefont {C.~M.}\ \bibnamefont
  {Marcus}},\ and\ \bibinfo {author} {\bibfnamefont {Y.}~\bibnamefont {Oreg}},\
  }\href {https://doi.org/10.1038/s41578-018-0003-1} {\bibfield  {journal}
  {\bibinfo  {journal} {Nature Reviews Materials}\ }\textbf {\bibinfo {volume}
  {3}},\ \bibinfo {pages} {52} (\bibinfo {year} {2018})}\BibitemShut {NoStop}%
\bibitem [{\citenamefont {Mourik}\ \emph {et~al.}(2012)\citenamefont {Mourik},
  \citenamefont {Zuo}, \citenamefont {Frolov}, \citenamefont {Plissard},
  \citenamefont {Bakkers},\ and\ \citenamefont
  {Kouwenhoven}}]{Mourik_science2012}%
  \BibitemOpen
  \bibfield  {author} {\bibinfo {author} {\bibfnamefont {V.}~\bibnamefont
  {Mourik}}, \bibinfo {author} {\bibfnamefont {K.}~\bibnamefont {Zuo}},
  \bibinfo {author} {\bibfnamefont {S.~M.}\ \bibnamefont {Frolov}}, \bibinfo
  {author} {\bibfnamefont {S.~R.}\ \bibnamefont {Plissard}}, \bibinfo {author}
  {\bibfnamefont {E.~P. A.~M.}\ \bibnamefont {Bakkers}},\ and\ \bibinfo
  {author} {\bibfnamefont {L.~P.}\ \bibnamefont {Kouwenhoven}},\ }\href
  {https://doi.org/10.1126/science.1222360} {\bibfield  {journal} {\bibinfo
  {journal} {Science}\ }\textbf {\bibinfo {volume} {336}},\ \bibinfo {pages}
  {1003} (\bibinfo {year} {2012})}\BibitemShut {NoStop}%
\bibitem [{\citenamefont {Nichele}\ \emph {et~al.}(2017)\citenamefont
  {Nichele}, \citenamefont {Drachmann}, \citenamefont {Whiticar}, \citenamefont
  {O'Farrell}, \citenamefont {Suominen}, \citenamefont {Fornieri},
  \citenamefont {Wang}, \citenamefont {Gardner}, \citenamefont {Thomas},
  \citenamefont {Hatke}, \citenamefont {Krogstrup}, \citenamefont {Manfra},
  \citenamefont {Flensberg},\ and\ \citenamefont {Marcus}}]{Nichele_PRL2017}%
  \BibitemOpen
  \bibfield  {author} {\bibinfo {author} {\bibfnamefont {F.}~\bibnamefont
  {Nichele}}, \bibinfo {author} {\bibfnamefont {A.~C.~C.}\ \bibnamefont
  {Drachmann}}, \bibinfo {author} {\bibfnamefont {A.~M.}\ \bibnamefont
  {Whiticar}}, \bibinfo {author} {\bibfnamefont {E.~C.~T.}\ \bibnamefont
  {O'Farrell}}, \bibinfo {author} {\bibfnamefont {H.~J.}\ \bibnamefont
  {Suominen}}, \bibinfo {author} {\bibfnamefont {A.}~\bibnamefont {Fornieri}},
  \bibinfo {author} {\bibfnamefont {T.}~\bibnamefont {Wang}}, \bibinfo {author}
  {\bibfnamefont {G.~C.}\ \bibnamefont {Gardner}}, \bibinfo {author}
  {\bibfnamefont {C.}~\bibnamefont {Thomas}}, \bibinfo {author} {\bibfnamefont
  {A.~T.}\ \bibnamefont {Hatke}}, \bibinfo {author} {\bibfnamefont
  {P.}~\bibnamefont {Krogstrup}}, \bibinfo {author} {\bibfnamefont {M.~J.}\
  \bibnamefont {Manfra}}, \bibinfo {author} {\bibfnamefont {K.}~\bibnamefont
  {Flensberg}},\ and\ \bibinfo {author} {\bibfnamefont {C.~M.}\ \bibnamefont
  {Marcus}},\ }\href {https://doi.org/10.1103/PhysRevLett.119.136803}
  {\bibfield  {journal} {\bibinfo  {journal} {Phys. Rev. Lett.}\ }\textbf
  {\bibinfo {volume} {119}},\ \bibinfo {pages} {136803} (\bibinfo {year}
  {2017})}\BibitemShut {NoStop}%
\bibitem [{\citenamefont {Fornieri}\ \emph {et~al.}(2019)\citenamefont
  {Fornieri}, \citenamefont {Whiticar}, \citenamefont {Setiawan}, \citenamefont
  {Portol{\'e}s}, \citenamefont {Drachmann}, \citenamefont {Keselman},
  \citenamefont {Gronin}, \citenamefont {Thomas}, \citenamefont {Wang},
  \citenamefont {Kallaher}, \citenamefont {Gardner}, \citenamefont {Berg},
  \citenamefont {Manfra}, \citenamefont {Stern}, \citenamefont {Marcus},\ and\
  \citenamefont {Nichele}}]{Fornieri_Nat2019}%
  \BibitemOpen
  \bibfield  {author} {\bibinfo {author} {\bibfnamefont {A.}~\bibnamefont
  {Fornieri}}, \bibinfo {author} {\bibfnamefont {A.~M.}\ \bibnamefont
  {Whiticar}}, \bibinfo {author} {\bibfnamefont {F.}~\bibnamefont {Setiawan}},
  \bibinfo {author} {\bibfnamefont {E.}~\bibnamefont {Portol{\'e}s}}, \bibinfo
  {author} {\bibfnamefont {A.~C.~C.}\ \bibnamefont {Drachmann}}, \bibinfo
  {author} {\bibfnamefont {A.}~\bibnamefont {Keselman}}, \bibinfo {author}
  {\bibfnamefont {S.}~\bibnamefont {Gronin}}, \bibinfo {author} {\bibfnamefont
  {C.}~\bibnamefont {Thomas}}, \bibinfo {author} {\bibfnamefont
  {T.}~\bibnamefont {Wang}}, \bibinfo {author} {\bibfnamefont {R.}~\bibnamefont
  {Kallaher}}, \bibinfo {author} {\bibfnamefont {G.~C.}\ \bibnamefont
  {Gardner}}, \bibinfo {author} {\bibfnamefont {E.}~\bibnamefont {Berg}},
  \bibinfo {author} {\bibfnamefont {M.~J.}\ \bibnamefont {Manfra}}, \bibinfo
  {author} {\bibfnamefont {A.}~\bibnamefont {Stern}}, \bibinfo {author}
  {\bibfnamefont {C.~M.}\ \bibnamefont {Marcus}},\ and\ \bibinfo {author}
  {\bibfnamefont {F.}~\bibnamefont {Nichele}},\ }\href
  {https://doi.org/10.1038/s41586-019-1068-8} {\bibfield  {journal} {\bibinfo
  {journal} {Nature}\ }\textbf {\bibinfo {volume} {569}},\ \bibinfo {pages}
  {89} (\bibinfo {year} {2019})}\BibitemShut {NoStop}%
\bibitem [{\citenamefont {Banerjee}\ \emph
  {et~al.}(2023{\natexlab{a}})\citenamefont {Banerjee}, \citenamefont {Lesser},
  \citenamefont {Rahman}, \citenamefont {Wang}, \citenamefont {Li},
  \citenamefont {Kringh\o{}j}, \citenamefont {Whiticar}, \citenamefont
  {Drachmann}, \citenamefont {Thomas}, \citenamefont {Wang}, \citenamefont
  {Manfra}, \citenamefont {Berg}, \citenamefont {Oreg}, \citenamefont {Stern},\
  and\ \citenamefont {Marcus}}]{Banerjee_PRB2023}%
  \BibitemOpen
  \bibfield  {author} {\bibinfo {author} {\bibfnamefont {A.}~\bibnamefont
  {Banerjee}}, \bibinfo {author} {\bibfnamefont {O.}~\bibnamefont {Lesser}},
  \bibinfo {author} {\bibfnamefont {M.~A.}\ \bibnamefont {Rahman}}, \bibinfo
  {author} {\bibfnamefont {H.-R.}\ \bibnamefont {Wang}}, \bibinfo {author}
  {\bibfnamefont {M.-R.}\ \bibnamefont {Li}}, \bibinfo {author} {\bibfnamefont
  {A.}~\bibnamefont {Kringh\o{}j}}, \bibinfo {author} {\bibfnamefont {A.~M.}\
  \bibnamefont {Whiticar}}, \bibinfo {author} {\bibfnamefont {A.~C.~C.}\
  \bibnamefont {Drachmann}}, \bibinfo {author} {\bibfnamefont {C.}~\bibnamefont
  {Thomas}}, \bibinfo {author} {\bibfnamefont {T.}~\bibnamefont {Wang}},
  \bibinfo {author} {\bibfnamefont {M.~J.}\ \bibnamefont {Manfra}}, \bibinfo
  {author} {\bibfnamefont {E.}~\bibnamefont {Berg}}, \bibinfo {author}
  {\bibfnamefont {Y.}~\bibnamefont {Oreg}}, \bibinfo {author} {\bibfnamefont
  {A.}~\bibnamefont {Stern}},\ and\ \bibinfo {author} {\bibfnamefont {C.~M.}\
  \bibnamefont {Marcus}},\ }\href {https://doi.org/10.1103/PhysRevB.107.245304}
  {\bibfield  {journal} {\bibinfo  {journal} {Phys. Rev. B}\ }\textbf {\bibinfo
  {volume} {107}},\ \bibinfo {pages} {245304} (\bibinfo {year}
  {2023}{\natexlab{a}})}\BibitemShut {NoStop}%
\bibitem [{\citenamefont {{Aghaee et al.}}(2023)}]{Aghaee_PRB2023_short}%
  \BibitemOpen
  \bibfield  {author} {\bibinfo {author} {\bibfnamefont {M.}~\bibnamefont
  {{Aghaee et al.}}} (\bibinfo {collaboration} {Microsoft Quantum}),\ }\href
  {https://doi.org/10.1103/PhysRevB.107.245423} {\bibfield  {journal} {\bibinfo
   {journal} {Phys. Rev. B}\ }\textbf {\bibinfo {volume} {107}},\ \bibinfo
  {pages} {245423} (\bibinfo {year} {2023})}\BibitemShut {NoStop}%
\bibitem [{\citenamefont {Banerjee}\ \emph
  {et~al.}(2023{\natexlab{b}})\citenamefont {Banerjee}, \citenamefont {Lesser},
  \citenamefont {Rahman}, \citenamefont {Thomas}, \citenamefont {Wang},
  \citenamefont {Manfra}, \citenamefont {Berg}, \citenamefont {Oreg},
  \citenamefont {Stern},\ and\ \citenamefont {Marcus}}]{Banerjee_PRL2023}%
  \BibitemOpen
  \bibfield  {author} {\bibinfo {author} {\bibfnamefont {A.}~\bibnamefont
  {Banerjee}}, \bibinfo {author} {\bibfnamefont {O.}~\bibnamefont {Lesser}},
  \bibinfo {author} {\bibfnamefont {M.~A.}\ \bibnamefont {Rahman}}, \bibinfo
  {author} {\bibfnamefont {C.}~\bibnamefont {Thomas}}, \bibinfo {author}
  {\bibfnamefont {T.}~\bibnamefont {Wang}}, \bibinfo {author} {\bibfnamefont
  {M.~J.}\ \bibnamefont {Manfra}}, \bibinfo {author} {\bibfnamefont
  {E.}~\bibnamefont {Berg}}, \bibinfo {author} {\bibfnamefont {Y.}~\bibnamefont
  {Oreg}}, \bibinfo {author} {\bibfnamefont {A.}~\bibnamefont {Stern}},\ and\
  \bibinfo {author} {\bibfnamefont {C.~M.}\ \bibnamefont {Marcus}},\ }\href
  {https://doi.org/10.1103/PhysRevLett.130.096202} {\bibfield  {journal}
  {\bibinfo  {journal} {Phys. Rev. Lett.}\ }\textbf {\bibinfo {volume} {130}},\
  \bibinfo {pages} {096202} (\bibinfo {year} {2023}{\natexlab{b}})}\BibitemShut
  {NoStop}%
\bibitem [{\citenamefont {Albrecht}\ \emph {et~al.}(2016)\citenamefont
  {Albrecht}, \citenamefont {Higginbotham}, \citenamefont {Madsen},
  \citenamefont {Kuemmeth}, \citenamefont {Jespersen}, \citenamefont
  {Nyg{\aa}rd}, \citenamefont {Krogstrup},\ and\ \citenamefont
  {Marcus}}]{Albrecht_Nature2016}%
  \BibitemOpen
  \bibfield  {author} {\bibinfo {author} {\bibfnamefont {S.~M.}\ \bibnamefont
  {Albrecht}}, \bibinfo {author} {\bibfnamefont {A.~P.}\ \bibnamefont
  {Higginbotham}}, \bibinfo {author} {\bibfnamefont {M.}~\bibnamefont
  {Madsen}}, \bibinfo {author} {\bibfnamefont {F.}~\bibnamefont {Kuemmeth}},
  \bibinfo {author} {\bibfnamefont {T.~S.}\ \bibnamefont {Jespersen}}, \bibinfo
  {author} {\bibfnamefont {J.}~\bibnamefont {Nyg{\aa}rd}}, \bibinfo {author}
  {\bibfnamefont {P.}~\bibnamefont {Krogstrup}},\ and\ \bibinfo {author}
  {\bibfnamefont {C.~M.}\ \bibnamefont {Marcus}},\ }\href
  {https://doi.org/10.1038/nature17162} {\bibfield  {journal} {\bibinfo
  {journal} {Nature}\ }\textbf {\bibinfo {volume} {531}},\ \bibinfo {pages}
  {206} (\bibinfo {year} {2016})}\BibitemShut {NoStop}%
\bibitem [{\citenamefont {Prada}\ \emph {et~al.}(2012)\citenamefont {Prada},
  \citenamefont {San-Jose},\ and\ \citenamefont {Aguado}}]{Prada_PRB2012}%
  \BibitemOpen
  \bibfield  {author} {\bibinfo {author} {\bibfnamefont {E.}~\bibnamefont
  {Prada}}, \bibinfo {author} {\bibfnamefont {P.}~\bibnamefont {San-Jose}},\
  and\ \bibinfo {author} {\bibfnamefont {R.}~\bibnamefont {Aguado}},\ }\href
  {https://doi.org/10.1103/PhysRevB.86.180503} {\bibfield  {journal} {\bibinfo
  {journal} {Phys. Rev. B}\ }\textbf {\bibinfo {volume} {86}},\ \bibinfo
  {pages} {180503} (\bibinfo {year} {2012})}\BibitemShut {NoStop}%
\bibitem [{\citenamefont {Kells}\ \emph {et~al.}(2012)\citenamefont {Kells},
  \citenamefont {Meidan},\ and\ \citenamefont {Brouwer}}]{Kells_PRB12}%
  \BibitemOpen
  \bibfield  {author} {\bibinfo {author} {\bibfnamefont {G.}~\bibnamefont
  {Kells}}, \bibinfo {author} {\bibfnamefont {D.}~\bibnamefont {Meidan}},\ and\
  \bibinfo {author} {\bibfnamefont {P.~W.}\ \bibnamefont {Brouwer}},\ }\href
  {https://doi.org/10.1103/PhysRevB.86.100503} {\bibfield  {journal} {\bibinfo
  {journal} {Phys. Rev. B}\ }\textbf {\bibinfo {volume} {86}},\ \bibinfo
  {pages} {100503} (\bibinfo {year} {2012})}\BibitemShut {NoStop}%
\bibitem [{\citenamefont {Liu}\ \emph {et~al.}(2012)\citenamefont {Liu},
  \citenamefont {Potter}, \citenamefont {Law},\ and\ \citenamefont
  {Lee}}]{Liu2012}%
  \BibitemOpen
  \bibfield  {author} {\bibinfo {author} {\bibfnamefont {J.}~\bibnamefont
  {Liu}}, \bibinfo {author} {\bibfnamefont {A.~C.}\ \bibnamefont {Potter}},
  \bibinfo {author} {\bibfnamefont {K.~T.}\ \bibnamefont {Law}},\ and\ \bibinfo
  {author} {\bibfnamefont {P.~A.}\ \bibnamefont {Lee}},\ }\href
  {https://doi.org/10.1103/physrevlett.109.267002} {\bibfield  {journal}
  {\bibinfo  {journal} {Phys. Rev. Lett.}\ }\textbf {\bibinfo {volume} {109}},\
  \bibinfo {pages} {267002} (\bibinfo {year} {2012})}\BibitemShut {NoStop}%
\bibitem [{\citenamefont {Liu}\ \emph {et~al.}(2017)\citenamefont {Liu},
  \citenamefont {Sau}, \citenamefont {Stanescu},\ and\ \citenamefont
  {Das~Sarma}}]{Liu2017}%
  \BibitemOpen
  \bibfield  {author} {\bibinfo {author} {\bibfnamefont {C.-X.}\ \bibnamefont
  {Liu}}, \bibinfo {author} {\bibfnamefont {J.~D.}\ \bibnamefont {Sau}},
  \bibinfo {author} {\bibfnamefont {T.~D.}\ \bibnamefont {Stanescu}},\ and\
  \bibinfo {author} {\bibfnamefont {S.}~\bibnamefont {Das~Sarma}},\ }\href
  {https://doi.org/10.1103/physrevb.96.075161} {\bibfield  {journal} {\bibinfo
  {journal} {Phys. Rev. B}\ }\textbf {\bibinfo {volume} {96}},\ \bibinfo
  {pages} {075161} (\bibinfo {year} {2017})}\BibitemShut {NoStop}%
\bibitem [{\citenamefont {Moore}\ \emph {et~al.}(2018)\citenamefont {Moore},
  \citenamefont {Zeng}, \citenamefont {Stanescu},\ and\ \citenamefont
  {Tewari}}]{Moore_PRB18}%
  \BibitemOpen
  \bibfield  {author} {\bibinfo {author} {\bibfnamefont {C.}~\bibnamefont
  {Moore}}, \bibinfo {author} {\bibfnamefont {C.}~\bibnamefont {Zeng}},
  \bibinfo {author} {\bibfnamefont {T.~D.}\ \bibnamefont {Stanescu}},\ and\
  \bibinfo {author} {\bibfnamefont {S.}~\bibnamefont {Tewari}},\ }\href
  {https://doi.org/10.1103/PhysRevB.98.155314} {\bibfield  {journal} {\bibinfo
  {journal} {Phys. Rev. B}\ }\textbf {\bibinfo {volume} {98}},\ \bibinfo
  {pages} {155314} (\bibinfo {year} {2018})}\BibitemShut {NoStop}%
\bibitem [{\citenamefont {Reeg}\ \emph {et~al.}(2018)\citenamefont {Reeg},
  \citenamefont {Dmytruk}, \citenamefont {Chevallier}, \citenamefont {Loss},\
  and\ \citenamefont {Klinovaja}}]{reeg2018zero}%
  \BibitemOpen
  \bibfield  {author} {\bibinfo {author} {\bibfnamefont {C.}~\bibnamefont
  {Reeg}}, \bibinfo {author} {\bibfnamefont {O.}~\bibnamefont {Dmytruk}},
  \bibinfo {author} {\bibfnamefont {D.}~\bibnamefont {Chevallier}}, \bibinfo
  {author} {\bibfnamefont {D.}~\bibnamefont {Loss}},\ and\ \bibinfo {author}
  {\bibfnamefont {J.}~\bibnamefont {Klinovaja}},\ }\href
  {https://doi.org/10.1103/PhysRevB.98.245407} {\bibfield  {journal} {\bibinfo
  {journal} {Phys. Rev. B}\ }\textbf {\bibinfo {volume} {98}},\ \bibinfo
  {pages} {245407} (\bibinfo {year} {2018})}\BibitemShut {NoStop}%
\bibitem [{\citenamefont {Awoga}\ \emph {et~al.}(2019)\citenamefont {Awoga},
  \citenamefont {Cayao},\ and\ \citenamefont {Black-Schaffer}}]{Awoga_PRL2019}%
  \BibitemOpen
  \bibfield  {author} {\bibinfo {author} {\bibfnamefont {O.~A.}\ \bibnamefont
  {Awoga}}, \bibinfo {author} {\bibfnamefont {J.}~\bibnamefont {Cayao}},\ and\
  \bibinfo {author} {\bibfnamefont {A.~M.}\ \bibnamefont {Black-Schaffer}},\
  }\href {https://doi.org/10.1103/PhysRevLett.123.117001} {\bibfield  {journal}
  {\bibinfo  {journal} {Phys. Rev. Lett.}\ }\textbf {\bibinfo {volume} {123}},\
  \bibinfo {pages} {117001} (\bibinfo {year} {2019})}\BibitemShut {NoStop}%
\bibitem [{\citenamefont {Vuik}\ \emph {et~al.}(2019)\citenamefont {Vuik},
  \citenamefont {Nijholt}, \citenamefont {Akhmerov},\ and\ \citenamefont
  {Wimmer}}]{Vuik_SciPost19}%
  \BibitemOpen
  \bibfield  {author} {\bibinfo {author} {\bibfnamefont {A.}~\bibnamefont
  {Vuik}}, \bibinfo {author} {\bibfnamefont {B.}~\bibnamefont {Nijholt}},
  \bibinfo {author} {\bibfnamefont {A.~R.}\ \bibnamefont {Akhmerov}},\ and\
  \bibinfo {author} {\bibfnamefont {M.}~\bibnamefont {Wimmer}},\ }\href
  {https://doi.org/10.21468/SciPostPhys.7.5.061} {\bibfield  {journal}
  {\bibinfo  {journal} {SciPost Phys.}\ }\textbf {\bibinfo {volume} {7}},\
  \bibinfo {pages} {61} (\bibinfo {year} {2019})}\BibitemShut {NoStop}%
\bibitem [{\citenamefont {Pan}\ and\ \citenamefont
  {Das~Sarma}(2020)}]{Pan_PRR20}%
  \BibitemOpen
  \bibfield  {author} {\bibinfo {author} {\bibfnamefont {H.}~\bibnamefont
  {Pan}}\ and\ \bibinfo {author} {\bibfnamefont {S.}~\bibnamefont
  {Das~Sarma}},\ }\href {https://doi.org/10.1103/PhysRevResearch.2.013377}
  {\bibfield  {journal} {\bibinfo  {journal} {Phys. Rev. Research}\ }\textbf
  {\bibinfo {volume} {2}},\ \bibinfo {pages} {013377} (\bibinfo {year}
  {2020})}\BibitemShut {NoStop}%
\bibitem [{\citenamefont {Prada}\ \emph {et~al.}(2020)\citenamefont {Prada},
  \citenamefont {San-Jose}, \citenamefont {de~Moor}, \citenamefont {Geresdi},
  \citenamefont {Lee}, \citenamefont {Klinovaja}, \citenamefont {Loss},
  \citenamefont {Nyg{\aa}rd}, \citenamefont {Aguado},\ and\ \citenamefont
  {Kouwenhoven}}]{Prada_review}%
  \BibitemOpen
  \bibfield  {author} {\bibinfo {author} {\bibfnamefont {E.}~\bibnamefont
  {Prada}}, \bibinfo {author} {\bibfnamefont {P.}~\bibnamefont {San-Jose}},
  \bibinfo {author} {\bibfnamefont {M.~W.~A.}\ \bibnamefont {de~Moor}},
  \bibinfo {author} {\bibfnamefont {A.}~\bibnamefont {Geresdi}}, \bibinfo
  {author} {\bibfnamefont {E.~J.~H.}\ \bibnamefont {Lee}}, \bibinfo {author}
  {\bibfnamefont {J.}~\bibnamefont {Klinovaja}}, \bibinfo {author}
  {\bibfnamefont {D.}~\bibnamefont {Loss}}, \bibinfo {author} {\bibfnamefont
  {J.}~\bibnamefont {Nyg{\aa}rd}}, \bibinfo {author} {\bibfnamefont
  {R.}~\bibnamefont {Aguado}},\ and\ \bibinfo {author} {\bibfnamefont {L.~P.}\
  \bibnamefont {Kouwenhoven}},\ }\href
  {https://doi.org/10.1038/s42254-020-0228-y} {\bibfield  {journal} {\bibinfo
  {journal} {Nat. Rev. Phys.}\ }\textbf {\bibinfo {volume} {2}},\ \bibinfo
  {pages} {575} (\bibinfo {year} {2020})}\BibitemShut {NoStop}%
\bibitem [{\citenamefont {Hess}\ \emph {et~al.}(2021)\citenamefont {Hess},
  \citenamefont {Legg}, \citenamefont {Loss},\ and\ \citenamefont
  {Klinovaja}}]{hess2021local}%
  \BibitemOpen
  \bibfield  {author} {\bibinfo {author} {\bibfnamefont {R.}~\bibnamefont
  {Hess}}, \bibinfo {author} {\bibfnamefont {H.~F.}\ \bibnamefont {Legg}},
  \bibinfo {author} {\bibfnamefont {D.}~\bibnamefont {Loss}},\ and\ \bibinfo
  {author} {\bibfnamefont {J.}~\bibnamefont {Klinovaja}},\ }\href
  {https://doi.org/10.1103/PhysRevB.104.075405} {\bibfield  {journal} {\bibinfo
   {journal} {Phys. Rev. B}\ }\textbf {\bibinfo {volume} {104}},\ \bibinfo
  {pages} {075405} (\bibinfo {year} {2021})}\BibitemShut {NoStop}%
\bibitem [{\citenamefont {Sau}\ and\ \citenamefont
  {Sarma}(2012)}]{Sau_NatComm2012}%
  \BibitemOpen
  \bibfield  {author} {\bibinfo {author} {\bibfnamefont {J.~D.}\ \bibnamefont
  {Sau}}\ and\ \bibinfo {author} {\bibfnamefont {S.~D.}\ \bibnamefont
  {Sarma}},\ }\href {https://doi.org/10.1038/ncomms1966} {\bibfield  {journal}
  {\bibinfo  {journal} {Nature Commun.}\ }\textbf {\bibinfo {volume} {3}},\
  \bibinfo {pages} {964} (\bibinfo {year} {2012})}\BibitemShut {NoStop}%
\bibitem [{\citenamefont {Recher}\ \emph {et~al.}(2001)\citenamefont {Recher},
  \citenamefont {Sukhorukov},\ and\ \citenamefont {Loss}}]{Recher_PRB2001}%
  \BibitemOpen
  \bibfield  {author} {\bibinfo {author} {\bibfnamefont {P.}~\bibnamefont
  {Recher}}, \bibinfo {author} {\bibfnamefont {E.~V.}\ \bibnamefont
  {Sukhorukov}},\ and\ \bibinfo {author} {\bibfnamefont {D.}~\bibnamefont
  {Loss}},\ }\href {https://doi.org/10.1103/PhysRevB.63.165314} {\bibfield
  {journal} {\bibinfo  {journal} {Phys. Rev. B}\ }\textbf {\bibinfo {volume}
  {63}},\ \bibinfo {pages} {165314} (\bibinfo {year} {2001})}\BibitemShut
  {NoStop}%
\bibitem [{\citenamefont {Hofstetter}\ \emph {et~al.}(2009)\citenamefont
  {Hofstetter}, \citenamefont {Csonka}, \citenamefont {Nyg{\aa}rd},\ and\
  \citenamefont {Sch{\"o}nenberger}}]{Hofstetter_Nature2009}%
  \BibitemOpen
  \bibfield  {author} {\bibinfo {author} {\bibfnamefont {L.}~\bibnamefont
  {Hofstetter}}, \bibinfo {author} {\bibfnamefont {S.}~\bibnamefont {Csonka}},
  \bibinfo {author} {\bibfnamefont {J.}~\bibnamefont {Nyg{\aa}rd}},\ and\
  \bibinfo {author} {\bibfnamefont {C.}~\bibnamefont {Sch{\"o}nenberger}},\
  }\href {https://doi.org/10.1038/nature08432} {\bibfield  {journal} {\bibinfo
  {journal} {Nature}\ }\textbf {\bibinfo {volume} {461}},\ \bibinfo {pages}
  {960} (\bibinfo {year} {2009})}\BibitemShut {NoStop}%
\bibitem [{\citenamefont {Herrmann}\ \emph {et~al.}(2010)\citenamefont
  {Herrmann}, \citenamefont {Portier}, \citenamefont {Roche}, \citenamefont
  {Yeyati}, \citenamefont {Kontos},\ and\ \citenamefont
  {Strunk}}]{Herrmann_PRL2010}%
  \BibitemOpen
  \bibfield  {author} {\bibinfo {author} {\bibfnamefont {L.~G.}\ \bibnamefont
  {Herrmann}}, \bibinfo {author} {\bibfnamefont {F.}~\bibnamefont {Portier}},
  \bibinfo {author} {\bibfnamefont {P.}~\bibnamefont {Roche}}, \bibinfo
  {author} {\bibfnamefont {A.~L.}\ \bibnamefont {Yeyati}}, \bibinfo {author}
  {\bibfnamefont {T.}~\bibnamefont {Kontos}},\ and\ \bibinfo {author}
  {\bibfnamefont {C.}~\bibnamefont {Strunk}},\ }\href
  {https://doi.org/10.1103/PhysRevLett.104.026801} {\bibfield  {journal}
  {\bibinfo  {journal} {Phys. Rev. Lett.}\ }\textbf {\bibinfo {volume} {104}},\
  \bibinfo {pages} {026801} (\bibinfo {year} {2010})}\BibitemShut {NoStop}%
\bibitem [{\citenamefont {F\"ul\"op}\ \emph {et~al.}(2015)\citenamefont
  {F\"ul\"op}, \citenamefont {Dom\'{\i}nguez}, \citenamefont {d'Hollosy},
  \citenamefont {Baumgartner}, \citenamefont {Makk}, \citenamefont {Madsen},
  \citenamefont {Guzenko}, \citenamefont {Nyg\aa{}rd}, \citenamefont
  {Sch\"onenberger}, \citenamefont {Levy~Yeyati},\ and\ \citenamefont
  {Csonka}}]{Fulop_PRL2015}%
  \BibitemOpen
  \bibfield  {author} {\bibinfo {author} {\bibfnamefont {G.}~\bibnamefont
  {F\"ul\"op}}, \bibinfo {author} {\bibfnamefont {F.}~\bibnamefont
  {Dom\'{\i}nguez}}, \bibinfo {author} {\bibfnamefont {S.}~\bibnamefont
  {d'Hollosy}}, \bibinfo {author} {\bibfnamefont {A.}~\bibnamefont
  {Baumgartner}}, \bibinfo {author} {\bibfnamefont {P.}~\bibnamefont {Makk}},
  \bibinfo {author} {\bibfnamefont {M.~H.}\ \bibnamefont {Madsen}}, \bibinfo
  {author} {\bibfnamefont {V.~A.}\ \bibnamefont {Guzenko}}, \bibinfo {author}
  {\bibfnamefont {J.}~\bibnamefont {Nyg\aa{}rd}}, \bibinfo {author}
  {\bibfnamefont {C.}~\bibnamefont {Sch\"onenberger}}, \bibinfo {author}
  {\bibfnamefont {A.}~\bibnamefont {Levy~Yeyati}},\ and\ \bibinfo {author}
  {\bibfnamefont {S.}~\bibnamefont {Csonka}},\ }\href
  {https://doi.org/10.1103/PhysRevLett.115.227003} {\bibfield  {journal}
  {\bibinfo  {journal} {Phys. Rev. Lett.}\ }\textbf {\bibinfo {volume} {115}},\
  \bibinfo {pages} {227003} (\bibinfo {year} {2015})}\BibitemShut {NoStop}%
\bibitem [{\citenamefont {Wang}\ \emph {et~al.}(2022)\citenamefont {Wang},
  \citenamefont {Dvir}, \citenamefont {Mazur}, \citenamefont {Liu},
  \citenamefont {van Loo}, \citenamefont {ten Haaf}, \citenamefont {Bordin},
  \citenamefont {Gazibegovic}, \citenamefont {Badawy}, \citenamefont {Bakkers},
  \citenamefont {Wimmer},\ and\ \citenamefont {Kouwenhoven}}]{Wang_Nat2022}%
  \BibitemOpen
  \bibfield  {author} {\bibinfo {author} {\bibfnamefont {G.}~\bibnamefont
  {Wang}}, \bibinfo {author} {\bibfnamefont {T.}~\bibnamefont {Dvir}}, \bibinfo
  {author} {\bibfnamefont {G.~P.}\ \bibnamefont {Mazur}}, \bibinfo {author}
  {\bibfnamefont {C.-X.}\ \bibnamefont {Liu}}, \bibinfo {author} {\bibfnamefont
  {N.}~\bibnamefont {van Loo}}, \bibinfo {author} {\bibfnamefont {S.~L.~D.}\
  \bibnamefont {ten Haaf}}, \bibinfo {author} {\bibfnamefont {A.}~\bibnamefont
  {Bordin}}, \bibinfo {author} {\bibfnamefont {S.}~\bibnamefont {Gazibegovic}},
  \bibinfo {author} {\bibfnamefont {G.}~\bibnamefont {Badawy}}, \bibinfo
  {author} {\bibfnamefont {E.~P. A.~M.}\ \bibnamefont {Bakkers}}, \bibinfo
  {author} {\bibfnamefont {M.}~\bibnamefont {Wimmer}},\ and\ \bibinfo {author}
  {\bibfnamefont {L.~P.}\ \bibnamefont {Kouwenhoven}},\ }\href
  {https://doi.org/10.1038/s41586-022-05352-2} {\bibfield  {journal} {\bibinfo
  {journal} {Nature}\ }\textbf {\bibinfo {volume} {612}},\ \bibinfo {pages}
  {448} (\bibinfo {year} {2022})}\BibitemShut {NoStop}%
\bibitem [{\citenamefont {Bordoloi}\ \emph {et~al.}(2022)\citenamefont
  {Bordoloi}, \citenamefont {Zannier}, \citenamefont {Sorba}, \citenamefont
  {Sch{\"o}nenberger},\ and\ \citenamefont {Baumgartner}}]{Bordoloi_Nat2022}%
  \BibitemOpen
  \bibfield  {author} {\bibinfo {author} {\bibfnamefont {A.}~\bibnamefont
  {Bordoloi}}, \bibinfo {author} {\bibfnamefont {V.}~\bibnamefont {Zannier}},
  \bibinfo {author} {\bibfnamefont {L.}~\bibnamefont {Sorba}}, \bibinfo
  {author} {\bibfnamefont {C.}~\bibnamefont {Sch{\"o}nenberger}},\ and\
  \bibinfo {author} {\bibfnamefont {A.}~\bibnamefont {Baumgartner}},\ }\href
  {https://doi.org/10.1038/s41586-022-05436-z} {\bibfield  {journal} {\bibinfo
  {journal} {Nature}\ }\textbf {\bibinfo {volume} {612}},\ \bibinfo {pages}
  {454} (\bibinfo {year} {2022})}\BibitemShut {NoStop}%
\bibitem [{\citenamefont {Leijnse}\ and\ \citenamefont
  {Flensberg}(2012{\natexlab{b}})}]{Leijnse_PRB2012}%
  \BibitemOpen
  \bibfield  {author} {\bibinfo {author} {\bibfnamefont {M.}~\bibnamefont
  {Leijnse}}\ and\ \bibinfo {author} {\bibfnamefont {K.}~\bibnamefont
  {Flensberg}},\ }\href {https://doi.org/10.1103/PhysRevB.86.134528} {\bibfield
   {journal} {\bibinfo  {journal} {Phys. Rev. B}\ }\textbf {\bibinfo {volume}
  {86}},\ \bibinfo {pages} {134528} (\bibinfo {year}
  {2012}{\natexlab{b}})}\BibitemShut {NoStop}%
\bibitem [{\citenamefont {Tsintzis}\ \emph {et~al.}(2023)\citenamefont
  {Tsintzis}, \citenamefont {{Seoane Souto}}, \citenamefont {Flensberg},
  \citenamefont {Danon},\ and\ \citenamefont {Leijnse}}]{tsintzis2023roadmap}%
  \BibitemOpen
  \bibfield  {author} {\bibinfo {author} {\bibfnamefont {A.}~\bibnamefont
  {Tsintzis}}, \bibinfo {author} {\bibfnamefont {R.}~\bibnamefont {{Seoane
  Souto}}}, \bibinfo {author} {\bibfnamefont {K.}~\bibnamefont {Flensberg}},
  \bibinfo {author} {\bibfnamefont {J.}~\bibnamefont {Danon}},\ and\ \bibinfo
  {author} {\bibfnamefont {M.}~\bibnamefont {Leijnse}},\ }\href
  {https://arxiv.org/abs/2306.16289} {\bibfield  {journal} {\bibinfo  {journal}
  {arXiv:2306.16289}\ } (\bibinfo {year} {2023})}\BibitemShut {NoStop}%
\bibitem [{\citenamefont {Boross}\ and\ \citenamefont
  {P{\'a}lyi}(2023)}]{Boross_2023}%
  \BibitemOpen
  \bibfield  {author} {\bibinfo {author} {\bibfnamefont {P.}~\bibnamefont
  {Boross}}\ and\ \bibinfo {author} {\bibfnamefont {A.}~\bibnamefont
  {P{\'a}lyi}},\ }\href {https://arxiv.org/abs/2305.08464} {\bibfield
  {journal} {\bibinfo  {journal} {arXiv:2305.08464}\ } (\bibinfo {year}
  {2023})}\BibitemShut {NoStop}%
\bibitem [{\citenamefont {Liu}\ \emph {et~al.}(2022{\natexlab{a}})\citenamefont
  {Liu}, \citenamefont {Wang}, \citenamefont {Dvir},\ and\ \citenamefont
  {Wimmer}}]{Liu_PRL2022}%
  \BibitemOpen
  \bibfield  {author} {\bibinfo {author} {\bibfnamefont {C.-X.}\ \bibnamefont
  {Liu}}, \bibinfo {author} {\bibfnamefont {G.}~\bibnamefont {Wang}}, \bibinfo
  {author} {\bibfnamefont {T.}~\bibnamefont {Dvir}},\ and\ \bibinfo {author}
  {\bibfnamefont {M.}~\bibnamefont {Wimmer}},\ }\href
  {https://doi.org/10.1103/PhysRevLett.129.267701} {\bibfield  {journal}
  {\bibinfo  {journal} {Phys. Rev. Lett.}\ }\textbf {\bibinfo {volume} {129}},\
  \bibinfo {pages} {267701} (\bibinfo {year} {2022}{\natexlab{a}})}\BibitemShut
  {NoStop}%
\bibitem [{\citenamefont {Dvir}\ \emph {et~al.}(2023)\citenamefont {Dvir},
  \citenamefont {Wang}, \citenamefont {van Loo}, \citenamefont {Liu},
  \citenamefont {Mazur}, \citenamefont {Bordin}, \citenamefont {ten Haaf},
  \citenamefont {Wang}, \citenamefont {van Driel}, \citenamefont {Zatelli},
  \citenamefont {Li}, \citenamefont {Malinowski}, \citenamefont {Gazibegovic},
  \citenamefont {Badawy}, \citenamefont {Bakkers}, \citenamefont {Wimmer},\
  and\ \citenamefont {Kouwenhoven}}]{Dvir2023}%
  \BibitemOpen
  \bibfield  {author} {\bibinfo {author} {\bibfnamefont {T.}~\bibnamefont
  {Dvir}}, \bibinfo {author} {\bibfnamefont {G.}~\bibnamefont {Wang}}, \bibinfo
  {author} {\bibfnamefont {N.}~\bibnamefont {van Loo}}, \bibinfo {author}
  {\bibfnamefont {C.-X.}\ \bibnamefont {Liu}}, \bibinfo {author} {\bibfnamefont
  {G.~P.}\ \bibnamefont {Mazur}}, \bibinfo {author} {\bibfnamefont
  {A.}~\bibnamefont {Bordin}}, \bibinfo {author} {\bibfnamefont {S.~L.~D.}\
  \bibnamefont {ten Haaf}}, \bibinfo {author} {\bibfnamefont {J.-Y.}\
  \bibnamefont {Wang}}, \bibinfo {author} {\bibfnamefont {D.}~\bibnamefont {van
  Driel}}, \bibinfo {author} {\bibfnamefont {F.}~\bibnamefont {Zatelli}},
  \bibinfo {author} {\bibfnamefont {X.}~\bibnamefont {Li}}, \bibinfo {author}
  {\bibfnamefont {F.~K.}\ \bibnamefont {Malinowski}}, \bibinfo {author}
  {\bibfnamefont {S.}~\bibnamefont {Gazibegovic}}, \bibinfo {author}
  {\bibfnamefont {G.}~\bibnamefont {Badawy}}, \bibinfo {author} {\bibfnamefont
  {E.~P. A.~M.}\ \bibnamefont {Bakkers}}, \bibinfo {author} {\bibfnamefont
  {M.}~\bibnamefont {Wimmer}},\ and\ \bibinfo {author} {\bibfnamefont {L.~P.}\
  \bibnamefont {Kouwenhoven}},\ }\href
  {https://doi.org/10.1038/s41586-022-05585-1} {\bibfield  {journal} {\bibinfo
  {journal} {Nature}\ }\textbf {\bibinfo {volume} {614}},\ \bibinfo {pages}
  {445} (\bibinfo {year} {2023})}\BibitemShut {NoStop}%
\bibitem [{\citenamefont {Bordin}\ \emph {et~al.}(2023)\citenamefont {Bordin},
  \citenamefont {Li}, \citenamefont {van Driel}, \citenamefont {Wolff},
  \citenamefont {Wang}, \citenamefont {ten Haaf}, \citenamefont {Wang},
  \citenamefont {van Loo}, \citenamefont {Kouwenhoven},\ and\ \citenamefont
  {Dvir}}]{Bordin_arXiv2023}%
  \BibitemOpen
  \bibfield  {author} {\bibinfo {author} {\bibfnamefont {A.}~\bibnamefont
  {Bordin}}, \bibinfo {author} {\bibfnamefont {X.}~\bibnamefont {Li}}, \bibinfo
  {author} {\bibfnamefont {D.}~\bibnamefont {van Driel}}, \bibinfo {author}
  {\bibfnamefont {J.~C.}\ \bibnamefont {Wolff}}, \bibinfo {author}
  {\bibfnamefont {Q.}~\bibnamefont {Wang}}, \bibinfo {author} {\bibfnamefont
  {S.~L.~D.}\ \bibnamefont {ten Haaf}}, \bibinfo {author} {\bibfnamefont
  {G.}~\bibnamefont {Wang}}, \bibinfo {author} {\bibfnamefont {N.}~\bibnamefont
  {van Loo}}, \bibinfo {author} {\bibfnamefont {L.~P.}\ \bibnamefont
  {Kouwenhoven}},\ and\ \bibinfo {author} {\bibfnamefont {T.}~\bibnamefont
  {Dvir}},\ }\href {https://arxiv.org/abs/2306.07696} {\bibfield  {journal}
  {\bibinfo  {journal} {arXiv:2306.07696}\ } (\bibinfo {year}
  {2023})}\BibitemShut {NoStop}%
\bibitem [{\citenamefont {Tsintzis}\ \emph {et~al.}(2022)\citenamefont
  {Tsintzis}, \citenamefont {{Seoane Souto}},\ and\ \citenamefont
  {Leijnse}}]{Tsintzis2022}%
  \BibitemOpen
  \bibfield  {author} {\bibinfo {author} {\bibfnamefont {A.}~\bibnamefont
  {Tsintzis}}, \bibinfo {author} {\bibfnamefont {R.}~\bibnamefont {{Seoane
  Souto}}},\ and\ \bibinfo {author} {\bibfnamefont {M.}~\bibnamefont
  {Leijnse}},\ }\href {https://doi.org/10.1103/PhysRevB.106.L201404} {\bibfield
   {journal} {\bibinfo  {journal} {Phys. Rev. B}\ }\textbf {\bibinfo {volume}
  {106}},\ \bibinfo {pages} {L201404} (\bibinfo {year} {2022})}\BibitemShut
  {NoStop}%
\bibitem [{\citenamefont {Sedlmayr}\ and\ \citenamefont
  {Bena}(2015)}]{Sedlmayr2015}%
  \BibitemOpen
  \bibfield  {author} {\bibinfo {author} {\bibfnamefont {N.}~\bibnamefont
  {Sedlmayr}}\ and\ \bibinfo {author} {\bibfnamefont {C.}~\bibnamefont
  {Bena}},\ }\href
  {https://journals.aps.org/prb/abstract/10.1103/PhysRevB.92.115115} {\bibfield
   {journal} {\bibinfo  {journal} {Phys. Rev. B}\ }\textbf {\bibinfo {volume}
  {92}},\ \bibinfo {pages} {115115} (\bibinfo {year} {2015})}\BibitemShut
  {NoStop}%
\bibitem [{\citenamefont {Sedlmayr}\ \emph {et~al.}(2016)\citenamefont
  {Sedlmayr}, \citenamefont {Aguiar-Hualde},\ and\ \citenamefont
  {Bena}}]{Sedlmayr2016}%
  \BibitemOpen
  \bibfield  {author} {\bibinfo {author} {\bibfnamefont {N.}~\bibnamefont
  {Sedlmayr}}, \bibinfo {author} {\bibfnamefont {J.~M.}\ \bibnamefont
  {Aguiar-Hualde}},\ and\ \bibinfo {author} {\bibfnamefont {C.}~\bibnamefont
  {Bena}},\ }\href
  {https://journals.aps.org/prb/abstract/10.1103/PhysRevB.93.155425} {\bibfield
   {journal} {\bibinfo  {journal} {Phys. Rev. B}\ }\textbf {\bibinfo {volume}
  {93}},\ \bibinfo {pages} {155425} (\bibinfo {year} {2016})}\BibitemShut
  {NoStop}%
\bibitem [{\citenamefont {Aksenov}\ \emph {et~al.}(2020)\citenamefont
  {Aksenov}, \citenamefont {Zlotnikov},\ and\ \citenamefont
  {Shustin}}]{Aksenov2020}%
  \BibitemOpen
  \bibfield  {author} {\bibinfo {author} {\bibfnamefont {S.~V.}\ \bibnamefont
  {Aksenov}}, \bibinfo {author} {\bibfnamefont {A.~O.}\ \bibnamefont
  {Zlotnikov}},\ and\ \bibinfo {author} {\bibfnamefont {M.~S.}\ \bibnamefont
  {Shustin}},\ }\href
  {https://journals.aps.org/prb/abstract/10.1103/PhysRevB.101.125431}
  {\bibfield  {journal} {\bibinfo  {journal} {Phys. Rev. B}\ }\textbf {\bibinfo
  {volume} {101}},\ \bibinfo {pages} {125431} (\bibinfo {year}
  {2020})}\BibitemShut {NoStop}%
\bibitem [{\citenamefont {Liu}\ \emph {et~al.}(2022{\natexlab{b}})\citenamefont
  {Liu}, \citenamefont {Pan}, \citenamefont {Setiawan}, \citenamefont
  {Wimmer},\ and\ \citenamefont {Sau}}]{Liu2022_PMMfusion}%
  \BibitemOpen
  \bibfield  {author} {\bibinfo {author} {\bibfnamefont {C.-X.}\ \bibnamefont
  {Liu}}, \bibinfo {author} {\bibfnamefont {H.}~\bibnamefont {Pan}}, \bibinfo
  {author} {\bibfnamefont {F.}~\bibnamefont {Setiawan}}, \bibinfo {author}
  {\bibfnamefont {M.}~\bibnamefont {Wimmer}},\ and\ \bibinfo {author}
  {\bibfnamefont {J.~D.}\ \bibnamefont {Sau}},\ }\href
  {https://doi.org/10.48550/ARXIV.2212.01653} {\bibfield  {journal} {\bibinfo
  {journal} {arXiv:2212.01653}\ } (\bibinfo {year}
  {2022}{\natexlab{b}})}\BibitemShut {NoStop}%
\bibitem [{\citenamefont {Deng}\ \emph {et~al.}(2016)\citenamefont {Deng},
  \citenamefont {Vaitiekėnas}, \citenamefont {Hansen}, \citenamefont {Danon},
  \citenamefont {Leijnse}, \citenamefont {Flensberg}, \citenamefont {Nygård},
  \citenamefont {Krogstrup},\ and\ \citenamefont {Marcus}}]{deng2016majorana}%
  \BibitemOpen
  \bibfield  {author} {\bibinfo {author} {\bibfnamefont {M.-T.}\ \bibnamefont
  {Deng}}, \bibinfo {author} {\bibfnamefont {S.}~\bibnamefont {Vaitiekėnas}},
  \bibinfo {author} {\bibfnamefont {E.~B.}\ \bibnamefont {Hansen}}, \bibinfo
  {author} {\bibfnamefont {J.}~\bibnamefont {Danon}}, \bibinfo {author}
  {\bibfnamefont {M.}~\bibnamefont {Leijnse}}, \bibinfo {author} {\bibfnamefont
  {K.}~\bibnamefont {Flensberg}}, \bibinfo {author} {\bibfnamefont
  {J.}~\bibnamefont {Nygård}}, \bibinfo {author} {\bibfnamefont
  {P.}~\bibnamefont {Krogstrup}},\ and\ \bibinfo {author} {\bibfnamefont
  {C.~M.}\ \bibnamefont {Marcus}},\ }\href
  {https://doi.org/10.1126/science.aaf3961} {\bibfield  {journal} {\bibinfo
  {journal} {Science}\ }\textbf {\bibinfo {volume} {354}},\ \bibinfo {pages}
  {1557} (\bibinfo {year} {2016})}\BibitemShut {NoStop}%
\bibitem [{\citenamefont {Prada}\ \emph {et~al.}(2017)\citenamefont {Prada},
  \citenamefont {Aguado},\ and\ \citenamefont {San-Jose}}]{Prada_PRB2017}%
  \BibitemOpen
  \bibfield  {author} {\bibinfo {author} {\bibfnamefont {E.}~\bibnamefont
  {Prada}}, \bibinfo {author} {\bibfnamefont {R.}~\bibnamefont {Aguado}},\ and\
  \bibinfo {author} {\bibfnamefont {P.}~\bibnamefont {San-Jose}},\ }\href
  {https://doi.org/10.1103/PhysRevB.96.085418} {\bibfield  {journal} {\bibinfo
  {journal} {Phys. Rev. B}\ }\textbf {\bibinfo {volume} {96}},\ \bibinfo
  {pages} {085418} (\bibinfo {year} {2017})}\BibitemShut {NoStop}%
\bibitem [{\citenamefont {Clarke}(2017)}]{Clarke_PRB2017}%
  \BibitemOpen
  \bibfield  {author} {\bibinfo {author} {\bibfnamefont {D.~J.}\ \bibnamefont
  {Clarke}},\ }\href {https://doi.org/10.1103/PhysRevB.96.201109} {\bibfield
  {journal} {\bibinfo  {journal} {Phys. Rev. B}\ }\textbf {\bibinfo {volume}
  {96}},\ \bibinfo {pages} {201109} (\bibinfo {year} {2017})}\BibitemShut
  {NoStop}%
\bibitem [{\citenamefont {Deng}\ \emph {et~al.}(2018)\citenamefont {Deng},
  \citenamefont {Vaitiek\ifmmode~\dot{e}\else \.{e}\fi{}nas}, \citenamefont
  {Prada}, \citenamefont {San-Jose}, \citenamefont {Nyg\aa{}rd}, \citenamefont
  {Krogstrup}, \citenamefont {Aguado},\ and\ \citenamefont
  {Marcus}}]{deng2018nonlocality}%
  \BibitemOpen
  \bibfield  {author} {\bibinfo {author} {\bibfnamefont {M.-T.}\ \bibnamefont
  {Deng}}, \bibinfo {author} {\bibfnamefont {S.}~\bibnamefont
  {Vaitiek\ifmmode~\dot{e}\else \.{e}\fi{}nas}}, \bibinfo {author}
  {\bibfnamefont {E.}~\bibnamefont {Prada}}, \bibinfo {author} {\bibfnamefont
  {P.}~\bibnamefont {San-Jose}}, \bibinfo {author} {\bibfnamefont
  {J.}~\bibnamefont {Nyg\aa{}rd}}, \bibinfo {author} {\bibfnamefont
  {P.}~\bibnamefont {Krogstrup}}, \bibinfo {author} {\bibfnamefont
  {R.}~\bibnamefont {Aguado}},\ and\ \bibinfo {author} {\bibfnamefont {C.~M.}\
  \bibnamefont {Marcus}},\ }\href {https://doi.org/10.1103/PhysRevB.98.085125}
  {\bibfield  {journal} {\bibinfo  {journal} {Phys. Rev. B}\ }\textbf {\bibinfo
  {volume} {98}},\ \bibinfo {pages} {085125} (\bibinfo {year}
  {2018})}\BibitemShut {NoStop}%
\bibitem [{\citenamefont {Gruñeiro}\ \emph {et~al.}(2023)\citenamefont
  {Gruñeiro}, \citenamefont {Alvarado}, \citenamefont {Yeyati},\ and\
  \citenamefont {Arrachea}}]{gruñeiro2023transport}%
  \BibitemOpen
  \bibfield  {author} {\bibinfo {author} {\bibfnamefont {L.}~\bibnamefont
  {Gruñeiro}}, \bibinfo {author} {\bibfnamefont {M.}~\bibnamefont {Alvarado}},
  \bibinfo {author} {\bibfnamefont {A.~L.}\ \bibnamefont {Yeyati}},\ and\
  \bibinfo {author} {\bibfnamefont {L.}~\bibnamefont {Arrachea}},\ }\href
  {https://arxiv.org/abs/2305.02040} {\bibfield  {journal} {\bibinfo  {journal}
  {arXiv:2305.02040}\ } (\bibinfo {year} {2023})}\BibitemShut {NoStop}%
\bibitem [{Note1()}]{Note1}%
  \BibitemOpen
  \bibinfo {note} {The notions of ECT and CAR are no longer valid in the regime
  where the dot $S$ couples strongly to 1 and 2. Nevertheless, the system can
  then still host PMM sweet spots~\cite {Tsintzis2022}.}\BibitemShut {Stop}%
\bibitem [{\citenamefont {Stepanenko}\ \emph {et~al.}(2012)\citenamefont
  {Stepanenko}, \citenamefont {Rudner}, \citenamefont {Halperin},\ and\
  \citenamefont {Loss}}]{Stepanenko2012}%
  \BibitemOpen
  \bibfield  {author} {\bibinfo {author} {\bibfnamefont {D.}~\bibnamefont
  {Stepanenko}}, \bibinfo {author} {\bibfnamefont {M.}~\bibnamefont {Rudner}},
  \bibinfo {author} {\bibfnamefont {B.~I.}\ \bibnamefont {Halperin}},\ and\
  \bibinfo {author} {\bibfnamefont {D.}~\bibnamefont {Loss}},\ }\href
  {https://doi.org/10.1103/PhysRevB.85.075416} {\bibfield  {journal} {\bibinfo
  {journal} {Phys. Rev. B}\ }\textbf {\bibinfo {volume} {85}},\ \bibinfo
  {pages} {075416} (\bibinfo {year} {2012})}\BibitemShut {NoStop}%
\bibitem [{\citenamefont {Kiršanskas}\ \emph {et~al.}(2017)\citenamefont
  {Kiršanskas}, \citenamefont {Pedersen}, \citenamefont {Karlström},
  \citenamefont {Leijnse},\ and\ \citenamefont {Wacker}}]{Kirsanskas_CPC2017}%
  \BibitemOpen
  \bibfield  {author} {\bibinfo {author} {\bibfnamefont {G.}~\bibnamefont
  {Kiršanskas}}, \bibinfo {author} {\bibfnamefont {J.~N.}\ \bibnamefont
  {Pedersen}}, \bibinfo {author} {\bibfnamefont {O.}~\bibnamefont
  {Karlström}}, \bibinfo {author} {\bibfnamefont {M.}~\bibnamefont
  {Leijnse}},\ and\ \bibinfo {author} {\bibfnamefont {A.}~\bibnamefont
  {Wacker}},\ }\href
  {https://doi.org/https://doi.org/10.1016/j.cpc.2017.07.024} {\bibfield
  {journal} {\bibinfo  {journal} {Computer Physics Communications}\ }\textbf
  {\bibinfo {volume} {221}},\ \bibinfo {pages} {317} (\bibinfo {year}
  {2017})}\BibitemShut {NoStop}%
\bibitem [{\citenamefont {Danon}\ \emph {et~al.}(2020)\citenamefont {Danon},
  \citenamefont {Hellenes}, \citenamefont {Hansen}, \citenamefont {Casparis},
  \citenamefont {Higginbotham},\ and\ \citenamefont
  {Flensberg}}]{Danon_PRL2020}%
  \BibitemOpen
  \bibfield  {author} {\bibinfo {author} {\bibfnamefont {J.}~\bibnamefont
  {Danon}}, \bibinfo {author} {\bibfnamefont {A.~B.}\ \bibnamefont {Hellenes}},
  \bibinfo {author} {\bibfnamefont {E.~B.}\ \bibnamefont {Hansen}}, \bibinfo
  {author} {\bibfnamefont {L.}~\bibnamefont {Casparis}}, \bibinfo {author}
  {\bibfnamefont {A.~P.}\ \bibnamefont {Higginbotham}},\ and\ \bibinfo {author}
  {\bibfnamefont {K.}~\bibnamefont {Flensberg}},\ }\href
  {https://doi.org/10.1103/PhysRevLett.124.036801} {\bibfield  {journal}
  {\bibinfo  {journal} {Phys. Rev. Lett.}\ }\textbf {\bibinfo {volume} {124}},\
  \bibinfo {pages} {036801} (\bibinfo {year} {2020})}\BibitemShut {NoStop}%
\bibitem [{\citenamefont {Puglia}\ \emph {et~al.}(2021)\citenamefont {Puglia},
  \citenamefont {Martinez}, \citenamefont {M\'enard}, \citenamefont {P\"oschl},
  \citenamefont {Gronin}, \citenamefont {Gardner}, \citenamefont {Kallaher},
  \citenamefont {Manfra}, \citenamefont {Marcus}, \citenamefont
  {Higginbotham},\ and\ \citenamefont {Casparis}}]{Puglia_PRB2021}%
  \BibitemOpen
  \bibfield  {author} {\bibinfo {author} {\bibfnamefont {D.}~\bibnamefont
  {Puglia}}, \bibinfo {author} {\bibfnamefont {E.~A.}\ \bibnamefont
  {Martinez}}, \bibinfo {author} {\bibfnamefont {G.~C.}\ \bibnamefont
  {M\'enard}}, \bibinfo {author} {\bibfnamefont {A.}~\bibnamefont {P\"oschl}},
  \bibinfo {author} {\bibfnamefont {S.}~\bibnamefont {Gronin}}, \bibinfo
  {author} {\bibfnamefont {G.~C.}\ \bibnamefont {Gardner}}, \bibinfo {author}
  {\bibfnamefont {R.}~\bibnamefont {Kallaher}}, \bibinfo {author}
  {\bibfnamefont {M.~J.}\ \bibnamefont {Manfra}}, \bibinfo {author}
  {\bibfnamefont {C.~M.}\ \bibnamefont {Marcus}}, \bibinfo {author}
  {\bibfnamefont {A.~P.}\ \bibnamefont {Higginbotham}},\ and\ \bibinfo {author}
  {\bibfnamefont {L.}~\bibnamefont {Casparis}},\ }\href
  {https://doi.org/10.1103/PhysRevB.103.235201} {\bibfield  {journal} {\bibinfo
   {journal} {Phys. Rev. B}\ }\textbf {\bibinfo {volume} {103}},\ \bibinfo
  {pages} {235201} (\bibinfo {year} {2021})}\BibitemShut {NoStop}%
\end{thebibliography}%

\newpage

\section{Sweet spot parameters}\label{app:parameters}

Table \ref{table1} presents the gate configurations used for the different sweet spots presented in the main text, using the full model that includes finite Zeeman splitting and on-site Coulomb repulsion. We find these sweet spots by maximizing the MP with respect to the values of the on-site potentials of the dots while staying at the even--odd ground state degeneracy. We keep the tunnel rates in the PMM system symmetric, implying that all sweet spots occur for $\varepsilon_1=\varepsilon_2$.

\begin{table}[!h]
\centering
\begin{tabular}[t]{|l|c|c|c|c|c|c}
\hline
$E_Z$ & $\varepsilon_1$ & $\varepsilon_S$ & $\varepsilon_2$ & MP\\
\hline
0.15 & -0.30603 & -0.56442 & -0.30603 & 0.66080\\
0.2 & -0.29419 & -0.50301 & -0.29419 & 0.76204\\
0.25 & -0.28088 & -0.46116 & -0.28088 & 0.82205\\
0.3 & -0.26818 & -0.43133 & -0.26818 & 0.86044\\
0.4 & -0.24629 & -0.39260 & -0.24629 & 0.90556\\
0.75 & -0.19795 & -0.34183 & -0.19795 & 0.96076\\
1.5 & -0.15355 & -0.32894 & -0.15355 & 0.98559\\
3 & -0.12292 & -0.34221 & -0.12292 & 0.99486\\
\hline
\end{tabular}
\caption{Tuning parameters for the sweet spots for the different values of the Zeeman splitting on the outer PMM dots investigated in the main text.
All energies are in units of $\Delta$.}
\label{table1}
\end{table}%

\section{Non-local conductance}\label{app:nonlocal}

In recent years, the non-local conductance has emerged as an useful probe to determine the local BCS charges of bound states~\cite{Danon_PRL2020} and to measure the gap reopening after the topological transition in superconducting nanowires~\cite{Puglia_PRB2021}. In this Appendix, we present additional results comparing the local and non-local conductance through the dot--PMM system.

\begin{figure}[b!] \centering
\includegraphics[width=1\linewidth]{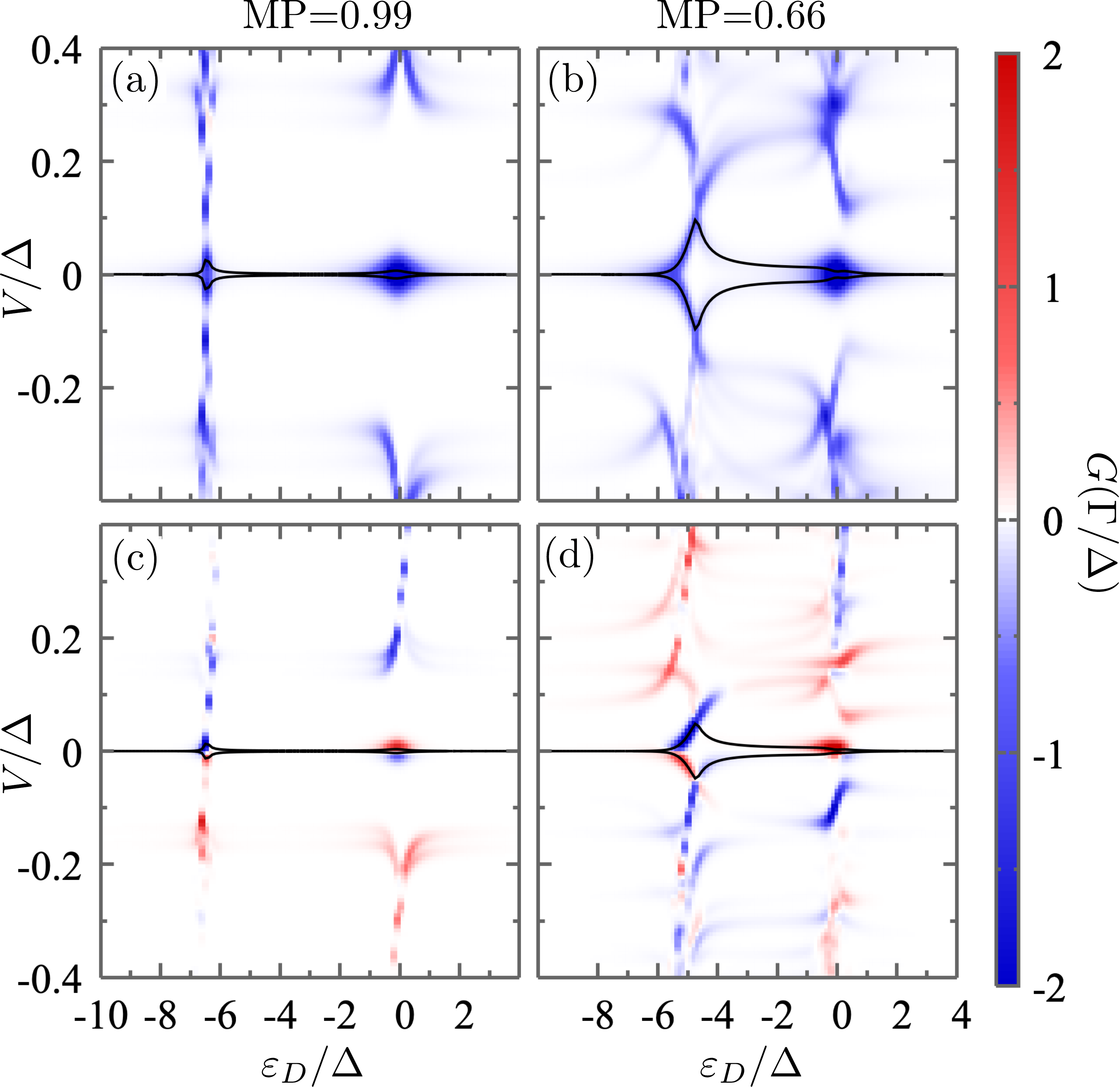}
\caption{(a,b) Local conductance at the left side, $G_{LL}$ and (c,d) non-local conductance, $G_{RL}$ through the dot--PMM system as a function of $\varepsilon_D$ and $V$. We included results for a high- (a,c) and a low-quality sweet spot (b,d). All parameters are the same as the ones used in Figs.~\ref{Fig:MP}(a,d). Features appear at twice the bias voltage in the top row, as the bias is applied symetrically to the system, in contrast to the lower row, where the bias is applied in the left lead, while keeping the right one at zero voltage.
}\label{Fig:nonLocal}
\end{figure}

We define the local and non-local conductance through
\begin{equation}
    G_{\alpha\beta}=\frac{I_{\alpha}(V_L,V_R)}{dV_\beta}\,,
\end{equation}
with $\alpha,\beta=L,R$ and $I_\alpha$ the current that enters the system through lead $\alpha$. For the local conductance, we assume that $V_L=-V_R=V/2$, while for the non-local conductance we use $V_\beta=V$ and $V_\alpha=0$.

In Fig.~\ref{Fig:nonLocal} we show the calculated $G_{LL}$ and $G_{RL}$ for a high- and a low-MP sweet spot, as a function of $\varepsilon_D$ and $V$.
[In Fig.~\ref{Fig:MP}(a,d) we showed the corresponding conductance through the right side of the system, $G_{RR}$.]
The conductance $G_{LL}$ through the left side [Fig.~\ref{Fig:nonLocal}(a,b)] has its maximum values close to $V=0$ for values of $\varepsilon_D$ where one of the spin-split levels on the additional dot is close to zero energy. Indeed, in this situation, resonant tunneling of electrons between the left lead and the additional dot is allowed for $V=0$.
The non-local conductance [Fig.~\ref{Fig:nonLocal}(c,d)] shows very similar features, with only few small differences.
For instance, $G_{LR}$ and $G_{RL}$ are identically zero close to zero bias for ideal sweet spots independently from $\varepsilon_D$ (results not shown).
(We obtain these ideal sweet spots from the full model by considering a very large Zeeman splitting in dots 1 and 2 of the PMM system.)
Deviations from the ideal Majorana case lead to an additional non-local conductance signal at low bias, which increases with decreasing MP, as can also be seen in Figs.~\ref{Fig:nonLocal}(c,d).

\end{document}